\documentclass{aa}  
\usepackage{graphicx}
\usepackage{txfonts}
\usepackage{lscape}
\usepackage{multirow}
\usepackage{footnote}
\usepackage{hyperref}  

\usepackage{multirow}
\usepackage{booktabs}

\usepackage{amsmath}
\usepackage{float}

\def\teff{$T_{\rm eff}$}
\def\logg{$\log g$}
\def\vsini{$v \sin i$}
\def\kms{km\,s$^{-1}$}

\begin{document}

   \title{The ESO UVES/FEROS Large Programs of TESS OB pulsators\thanks{Based on observations collected at the European Southern Observatory, Chile under programs 1104.D-0230 and 0104.A-9001}}

   \subtitle{II. On the physical origin of macroturbulence}

\author{Nadya Serebriakova\inst{\ref{KUL}}
\and Andrew Tkachenko\inst{\ref{KUL}}
\and Conny Aerts\inst{\ref{KUL},\ref{MPIA},\ref{Radboud}}
}
\institute{Institute of Astronomy, KU Leuven, Celestijnenlaan 200D, B-3001 Leuven, Belgium \\ \email{nadya.serebriakova@kuleuven.be} \label{KUL} 
\and Max Planck Institute for Astronomy, K\"onigstuhl 17, 69117 Heidelberg, Germany \label{MPIA}
\and Department of Astrophysics, IMAPP, Radboud University Nijmegen, PO Box 9010, 6500 GL Nijmegen, The Netherlands\label{Radboud}
}

   \date{Received ...; accepted ...}


  \abstract
   { Spectral lines of hot massive stars are known to exhibit large excess broadening in addition to rotational broadening. This excess broadening is often attributed to macroturbulence whose physical origin is a matter of active debate in the stellar astrophysics community.  }
   { We aim to shed light on the physical origin of macroturbulent line broadening by looking into the statistical properties of a large sample of O- and B-type stars, both in the Galaxy and The Large Magellanic Cloud (LMC).}
   { We deliver newly measured macroturbulent velocities for 86 stars from the Galaxy in a consistent manner with 126 stars from the LMC. A total sample of 594 stars with measured macroturbulent velocities was composed by complementing our sample with archival data for the Galactic O- and B-type stars for a better coverage of the parameter space. Furthermore, we compute an extensive grid of {\sc mesa} models to compare, in a statistical manner, the predicted interior properties of stars (such as convection and wave propagation) with the inference of macroturbulent velocities from high-resolution spectroscopic observations.}
   { We find evidence for subsurface convective zones formed in the iron opacity bump (FeCZ) being connected to observed macroturbulent velocities in hot massive stars. Additionally, we find the presence of two principally different regimes where, depending on the initial stellar mass, different mechanisms may be responsible for the observed excess line broadening. }
  { Stars with initial masses above 30$M_{\odot}$ exhibit macroturbulent velocities that are in line with FeCZ properties, as indicated by the trends in both observations and models. For stars below 12$M_{\odot}$, alternative mechanisms are needed to explain macroturbulent broadening, such as internal gravity waves (IGWs). Finally, in the intermediate range between 12 and 30$M_{\odot}$, IGWs tunnelling through subsurface convective layers, combined with the presence of FeCZ-driven convection, suggests that both processes could contribute to the observed macroturbulent velocities. This intermediate regime presents a region where the interplay between these two (or more) mechanisms remains to be fully understood.
  }

   \keywords{stars: fundamental parameters -- stars: massive -- stars: early-type -- stars: oscillations -- techniques: spectroscopic }

   \maketitle


\section{Introduction}\label{sec:intoduction}

Macroturbulent line broadening in stellar spectra refers to the broadening of spectral lines caused by large-scale turbulent motions in the stellar atmospheres. Historically, macroturbulent line broadening was identified as a key mechanism decades ago, when it was noticed that rotational broadening alone could not account for the observed widths and shapes of spectral lines. This led to the development of models incorporating large-scale atmospheric turbulence. A comprehensive review on the matter exists thanks to \citet{gray_turbulence_1978, Gray2008}, or, concerning massive stars, \citet{howarth_rotation_2004}. 

Particularly strong macroturbulent broadening was found in spectral lines of massive stars, characterised by macroturbulent velocities $\Theta$ ten times larger than those found in solar-type stars, often dominating over the rotational broadening. 

\citet{conti_ebbets_1977} analysed spectra of 205 O-type stars and compared rotational line broadening in main-sequence and evolved populations. The authors proposed the need for additional turbulent broadening to explain observed differences in distributions. The excess broadening was further quantified by \citet{howarth_cross_1997}, who 
measured the total line broadening in 373 O- and B- type supergiants in a self-consistent way by cross-correlating all observed spectra with one observed template spectrum. They reported the need for additional broadening mechanisms on top of the rotation to be able to explain the observed total line broadening of over 100 \kms~ for such a large sample, with presumably randomly distributed inclination angles. 

With the development of models for macroturbulent line profiles, it became possible to fit the total shapes of individual line profiles by convolution of intrinsic line profiles with both rotational and macroturbulent profiles.    \citet{ryans_macroturbulent_2002} analysed high-resolution spectra of 12 B-type supergiants and found that rotational broadening alone was not sufficient to describe observed line profile shapes, and extra macroturbulent broadening of 20-60 \kms~ had to be applied. Even higher macroturbulent velocities of up to 100 \kms~ were found by \cite{simon-diaz_observational_2010} in a sample of 15 supergiants of late-O to B spectral types. The same results were later reported in larger samples, such as the sample of 430 O- and B- type Galactic stars analysed by \citet{simon-diaz_iacob_2017} and the sample of 126 B-type supergiants from the LMC analysed by \citet{Serebriakova2023}.

The physical origin of such large velocities present in the  photospheres of hot massive stars is still debated. In the Sun, photospheric granulation is directly observed and leads to the macroturbulent broadening of lines in the solar spectrum (and spectra of other solar-type stars). Massive hot stars have radiative envelopes, hence, the macroturbulent broadening in these stars requires another physical explanation. Often it is attributed to the excitation of multiple non-radial pulsations modes (\citet{lucy_analysis_1976, kaufer_long-term_1997, ryans_macroturbulent_2002, howarth_rotation_2004}). 
\citet{kaufer_long-term_1997} performed a long-term monitoring of 6 BA supergiants and reported variability in radial velocities and line strengths with multiple-mode cyclic behaviour on time scales of 5 to 50 days. The authors attributed the detected variability to the effect of non-radial oscillations on the stellar surface. \citet{aerts_collective_2009,aerts_origin_2009} modelled spectral line profiles by the collective effect of hundreds of low-amplitude non-radial gravity modes on top of rotation. The authors reported that the resulting macroturbulent profile mimics these collective pulsational effects, and ignoring time-dependent waves leads to single snapshot line profiles with global velocity fields of unrealistic (sometimes supersonic) scale. 

While there seems to be agreement that macroturbulent velocities originate from pulsation modes and internal gravity waves (IGW), the debates are focused on where they get excited: at the interface of the convective core and the radiative envelope (\citet{rogers_internal_2013}) or subsurface convective layers. The subsurface convective layers originate in radiative envelopes of massive stars with an initial mass larger than some 10 $M_{\odot}$ due to the iron (FeCZ) or helium (HeCZ) opacity bumps. \citet{cantiello_sub-surface_2009} computed models of interiors of stars with masses from 5 to 100 $M_{\odot}$ and found correlations between the presence of FeCZ and observed microturbulent velocities $\xi$ and wind clumping.  
Furthermore, \citet{grassitelli_observational_2015, grassitelli_metallicity_2016} investigated FeCZ and found that areas of high turbulent pressure in these subsurface convective layers, when mapped to the positions in Hertzsprung-Russell diagram (HRD), aligned nicely with regions where observations showed an excess of stars with high macroturbulent velocity values. This result was reinforced by the conclusions of
\citet{simon-diaz_iacob_2017} who analysed macroturbulent velocities in a large sample of 430 massive stars. 

The debates were boosted with the discovery of stochastic low-frequency variability (SLFV) in space photometry of O- and B-type stars. The SLFV manifests itself as a power excess in the low-frequency part of the Fourier spectrum. While the first observational detections of SLFV date back to CoRoT and early $Kepler$ times \citep[e.g.,][]{Blomme2011,Tkachenko2014,Aerts2017a}, detailed observational studies and interpretation were first presented in \citet{bowman_photometric_2019, bowman_low-frequency_2019}. Furthermore, \citet{Burssens2020} and \citet{bowman_photometric_2020} found a correlation between SLFV properties and macroturbulent broadening, a finding suggestive of a connection between the two observed phenomena, both caused by the modes excited near the boundary of the convective core. 

\citet{lecoanet_low-frequency_2019, lecoanet_surface_2021} provided a theoretical model for internal gravity waves (IGWs) and concluded that the calculated surface manifestation does not correlate with the observed SLFV nor macroturbulence. At the same time, \citet{edelmann_three-dimensional_2019} and \citet{vanon_three-dimensional_2023} performed 3D simulations of core convection for 3 and 7$M_{\odot}$ stellar models, respectively, and reported theoretical frequency spectra to have properties similar to those of the observed SLFV. In particular, the authors noted a pronounced low-frequency excess present in the models with the frequency range decreasing with stellar age. Furthermore, \citet{ratnasingam_two-dimensional_2020} and \citet{ratnasingam_internal_2023} performed 2D simulations and computed frequency spectra of IGWs from 3, 5, 7, 10, and 13$M_{\odot}$ stellar models. The authors reported the slopes of the frequency spectra to be in agreement with observed properties of SLFV.
\citet{anders_photometric_2023} challenged higher masses and performed comprehensive simulations of core convection for 3, 15 and 40$M_{\odot}$ stellar models at the zero-age main sequence (ZAMS). From their simulations, the authors predicted the frequency spectra of the gravity waves propagated to the surface. Contrary to the previous findings based on 2- and 3-D simulations, \citet{anders_photometric_2023} concluded that all three models predict variability amplitudes several orders of magnitude lower than both observed SLFV amplitudes and current observational detection capabilities. 

Apart from IGWs, the subsurface convective layers were also simulated and tested as a plausible explanation of SLFV and macroturbulent broadening. \citet{cantiello_origin_2021} found that model properties of FeCZ from 1D  {\sc mesa} simulations in a large range of masses and ages correlate well with both SLFV and macroturbulence. A similar conclusion was presented in \citet{ma_variability_2024}. 3D simulations of exteriors were computed by \citet{schultz_stochastic_2022} and \citet{schultz_turbulence-supported_2023} for  a 13$M_{\odot}$ star at terminal age main sequence (TAMS), 35$M_{\odot}$ star at ZAMS and at mid-main sequence, and 56 and 80$M_{\odot}$ stars in the Hertzsprung gap. The authors concluded that FeCZ and HeCZ extend significantly further in radial direction compared to 1D models and reach stellar photosphere. The resulting large turbulent velocities at the stellar surface were found in agreement with similarly large observed macroturbulent velocities. Moreover, the same models were used by \citet{schultz_synthesizing_2023} to synthesise spectral line profiles broadened by the clumpy turbulent surface. The authors concluded on the similarity between the shapes of the macroturbulent broadening-dominated profile and their simulated line profiles. 

In this study, we present a statistical analysis of a sample of 594 stars for which observational inference of the macroturbulent velocity is available. The sample covers masses from 2.5 to 80 $M_{\odot}$ and two metallicity regimes (LMC and the Galaxy). For 86 stars, we present macroturbulent velocities measured for the first time. In addition to the observed sample, we compute a dense grid of {\sc mesa} stellar structure and evolution models and investigate the properties of FeCZ and HeCZ across the HRD. We compare the statistical properties of the observed sample and model properties of FeCZ in order to examine their possible connection. Moreover, for the full grid of models, we estimate the theoretical possibility of IGWs tunnelling through convective layers and assess its possible connection to the observations. 

The full description of observations and analyses involved in the sample is given in section \ref{sec:obsSamples}. Section \ref{sec:mesa} summarises input physics included in the computation of the {\sc mesa} models. Section \ref{sec:conv_vs_vmacro} is dedicated to qualitative and quantitative analyses of the observed macroturbulent velocities and model convective velocities, delving into multivariate linear regression in subsection \ref{sec:conv_vs_vmacro_multivar} and principal component analysis in subsection \ref{sec:conv_vs_vmacro_pca}. Finally, section \ref{sec:igw} explores IGWs propagation properties. The conclusions of our study are presented in section \ref{sec:discussion_conclusions}.

\section{Observed sample, data reduction, and spectroscopic parameter determination}\label{sec:obsSamples}

\begin{figure*}[!htbp]
   \begin{centering}
            {\includegraphics[clip,width=540pt,trim={1.0cm 1.0cm 0.0cm 1.0cm}]{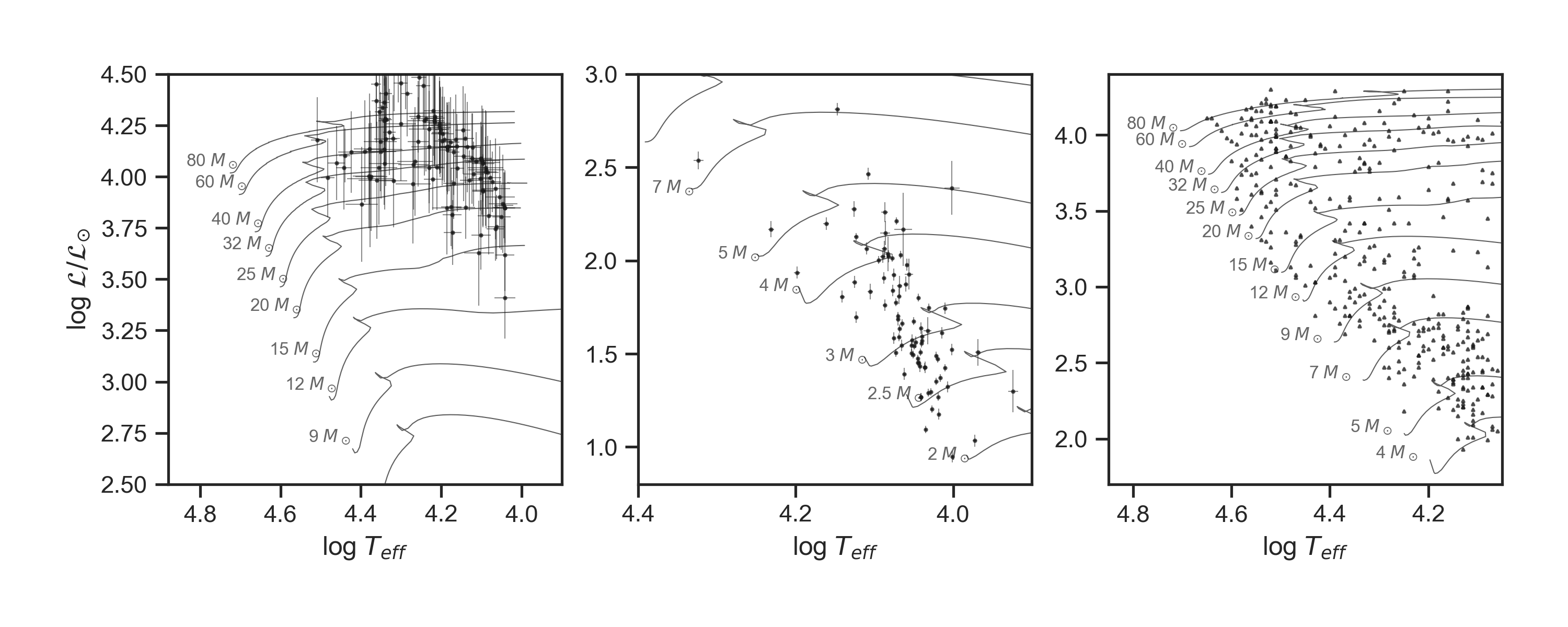}}
      \caption{Full observed sample of high-mass stars with measured macroturbulent velocities. Left panel: LMC supergiants from \citet{Serebriakova2023} over evolutionary tracks for $Z$=0.006. Middle and right panels: Galactic objects from \citet{Gebruers2022} and \citet{Simon-Diaz2017}, respectively, over evolutionary tracks for $Z$=0.014.}
         \label{fig:HRD1}
   \end{centering}
\end{figure*}

Our study focuses on a comprehensive sample of stars covering a mass range from 2.5 to 80 $M_{\odot}$ with measured macroturbulent velocities $\Theta$. The sample consists of data from two ESO Large Programs using the {\sc uves} and {\sc feros} spectrographs: 126 supergiants (SG) in the Large Magellanic Cloud (LMC) from \citet{Serebriakova2023}, 86 Galactic main-sequence (MS) stars from \citet{Gebruers2022}, and 382 Galactic MS and SG stars from \citet{Simon-Diaz2017}. This results in a total of 594 stars, covering metallicity regimes of the Milky Way (Solar metallicity of $Z$=0.014 is adopted for this regime) and LMC ($Z$=0.006) galaxies. The full sample is shown on the HRD in Fig.\ref{fig:HRD1}. 
Note that we plot the so-called spectroscopic luminosity defined by 
$\mathcal{L} \equiv T_{\rm eff}^4/g$ for the observed stars in all figures, following \citet{simon-diaz_iacob_2017}. This spectroscopic luminosity can be inferred directly from observed \teff\ and \logg\ values without requiring to know the distance 
and extinction, while it behaves similarly to the actual luminosity of the star \citep{langer_spectroscopic_2014}
and hence can be used as a proxy for it. The use of the observed spectroscopic luminosity in the spectroscopic HRD may introduce shifts of individual objects compared to their position in the classical HRD, thus affecting evolutionary mass determination. This is particularly the case  when uncertainties in 
\logg\ are high \citep[as seen in supergiants; see][]{markova2018, holgado2020}. However, it remains a valid tool for statistical analyses of large stellar samples as long as no conclusions are drawn about the individual stellar masses.

The data reduction, atmospheric parameter determination, and line broadening measurement for the three subsamples were handled primarily within their respective studies. Our work involved using the processed data and applying additional steps where necessary, specifically for the Galactic MS stars from \citet{Gebruers2022} where macroturbulent velocity measurements were not included in the analysis. 
The data reduction and processing varied slightly across the different subsamples due to the distinct methodologies employed in each of the referenced studies. Here, we provide a brief summary of the analyses employed for each subsample.

\subsection{Subsample of LMC supergiants from \citet{Serebriakova2023}}\label{sec:subsamp_serebriakova}

The observations were conducted as part of the large programs using the Ultraviolet and Visual Echelle Spectrograph ({\sc uves}; \citet{uves}) mounted on UT2 at the VLT and  the Fiber-fed Extended Range Optical Spectrograph ({\sc feros}; \citet{Kaufer1999_feros}) mounted on the MPG/ESO 2.2-m telescope. These programs aimed to deliver high-resolution, high signal-to-noise ratio spectroscopy of a sample of apparently single OB stars in the LMC. Each target was observed at two epochs separated by at least 2-3 weeks to detect potential spectral variability.
The data reduction process employed the  {\sc EsoReflex} data reduction pipeline for the  {\sc uves} spectra and the {\sc ceres} \citep{Brahm2017_ceres}  pipeline for the  {\sc feros} spectra. Custom order-by-order normalisation was applied to correct for artefacts resulting from the order merging of the original pipelines.

The atmospheric parameters were determined using a grid search method involving the  {\sc tlusty} \citep{hubeny1988_tlusty},  {\sc cmfgen} \citep{hiller2001_cmfgen}, and  {\sc gssp} \citep{tkachenko2015_gssp} model spectra, depending on the parameter range.
The projected rotational velocities \vsini~ were inferred using the Fourier transform (FT) method. This technique identifies the position of the first zero in the FT of a rotationally broadened profile, directly translating to the \vsini~ of the star. Specific spectral lines such as \ion{Si}{ii}, \ion{Si}{iii}, \ion{Mg}{ii}, and \ion{He}{i} were selected based on their suitability for different types of stars in the sample.
Macroturbulent velocities $\Theta$ were deduced by fitting synthetic spectra to observed line profiles of the \ion{Mg}{ii} 4481 Å line. The synthetic intrinsic profiles in the form of specific intensities $I(\mu)$ were computed for every object's given atmospheric parameters and convolved with a radial-tangential macroturbulent kernel. With the integration of $I(\mu)$ over the sphere, rotational broadening naturally included realistic limb darkening for the exact atmosphere model of each object. The projected rotational velocities were allowed to vary within 1$\sigma$ uncertainties of the values inferred with the FT method.

\subsection{Subsample of Galactic main-sequence stars from \citet{Gebruers2022}}\label{sec:subsamp_gebruers}

The observations were conducted as part of the same  {\sc feros} large program as in \citet{Serebriakova2023} albeit focused on Galactic main-sequence OB-type stars.
The modified  {\sc ceres} pipeline was employed for the extraction of stellar spectra. In the original study, the normalisation function was optimised together with atmospheric parameters. For the sake of consistency with \citet{Serebriakova2023}, we employed the same order-by-order approach to normalise the spectra.
Atmospheric parameters were derived using the  {\sc zeta-Payne} \citep{straumit2022_zetapayne} pipeline, which employs a pre-trained neural network to predict normalised synthetic spectra for a given set of stellar labels (\teff, \logg, \vsini, [M/H]). The normalisation function was optimised along with these labels.

Since \vsini~ derived by {\sc zeta-Payne} represents the total broadening due to the collective effect of rotation and macroturbulence (as the latter was not included in the fitting), we additionally employed the same method as in \citet{Serebriakova2023} to infer \vsini~ values, namely, using the FT method to separate the rotational broadening.
The original study did not include measurements of $\Theta$, hence we repeated the same procedure as in \citet{Serebriakova2023} to infer $\Theta$, which implied fitting the \ion{Mg}{ii} 4481 Å line profile with both rotationally and macroturbulent broadened synthetic intrinsic profiles.

\subsection{Subsample of Galactic main-sequence and supergiant stars from \citet{Simon-Diaz2017}}\label{sec:subsamp_simon-diaz}

The sample was observed as part of the  {\sc iacob} project \citep{diaz2011_iacob}, utilising high-resolution spectra from the Fiber-fed Echelle Spectrograph ({\sc fies}; \citet{telting2014_fies}) and the High-Efficiency and high-Resolution Mercator Echelle Spectrograph ({\sc hermes}; \citet{raskin2011_hermes})  attached to the 2.56-m Nordic Optical Telescope and 1.2-m Mercator Telescope, respectively.
The spectra were reduced using the {\sc FIEStool} and {\sc HermesDRS} pipelines for  {\sc fies} and  {\sc hermes}, respectively, followed by normalisation using custom procedures implemented in IDL.

The atmospheric parameters were obtained using the {\sc iacob-gbat} tool with a grid of {\sc fastwind} (\citet{puls2005_fastwind}) synthetic spectra.
Projected rotational velocities were determined using the  {\sc iacob-broad} tool, which combines the FT and goodness-of-fit (GOF) methodologies.
The same {\sc iacob-broad} tool was used to fit radial-tangential macroturbulent and rotational profiles. The FT method provided initial estimates, while the GOF method was used to refine the final values for both broadenings.

\subsection{A look into the full combined sample of the observed macroturbulent velocities $\Theta$}\label{sec:obs_vmacro}

\begin{figure*}[!htbp]
   \begin{centering}
            {\includegraphics[clip,width=540pt,trim={1.0cm 1.0cm 0.0cm 0.0cm}]{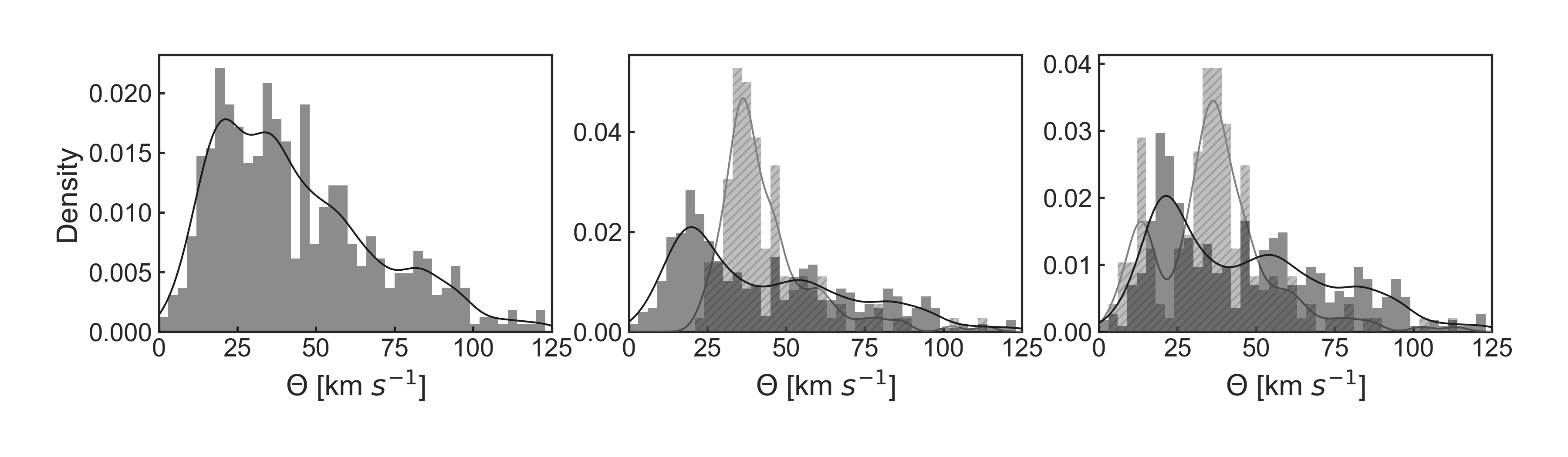}}
      \caption{Histograms and Kernel Density Estimates (KDE, solid line) of the measured $\Theta$. From left to right: the full sample of 594 stars, Galactic (dark grey) versus LMC (light grey) stars, and stars for which macroturbulent velocities were inferred with the methods of \citet[][light grey]{Serebriakova2023} and \citet[][dark grey]{simon-diaz_iacob_2017}.}
         \label{fig:theta_observed_hist}
   \end{centering}
\end{figure*}

\begin{figure*}
   \sidecaption
            {\includegraphics[clip,width=12cm,trim={1.0cm 1.0cm 0.0cm 0.0cm}]{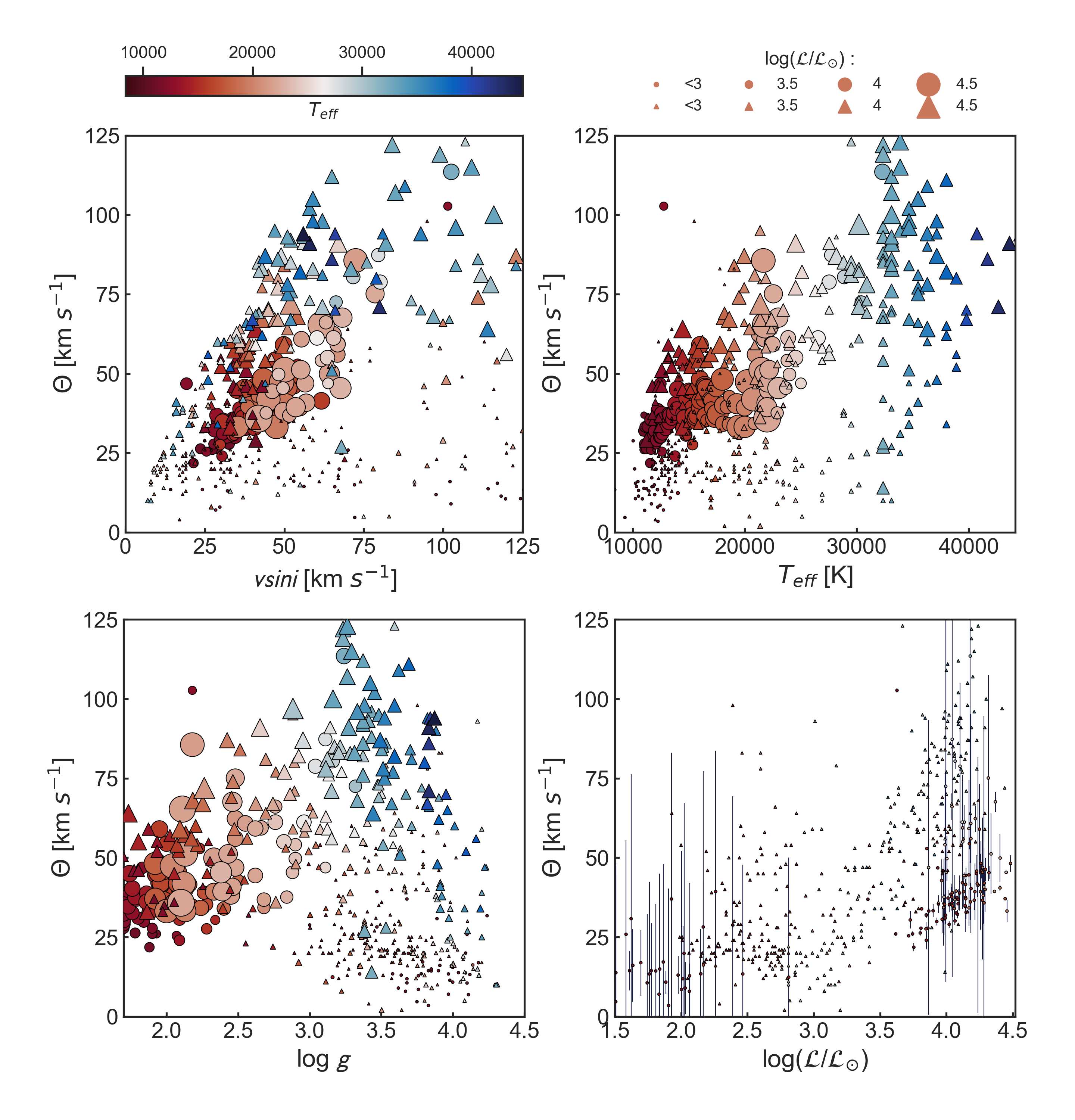}}
      \caption{Macroturbulent velocity as a function of stellar parameters. From top left to bottom right:$\Theta$ vs \vsini, $\Theta$ vs \teff, $\Theta$ vs \logg, and $\Theta$ vs $\log (\mathcal{L}/\mathcal{L_{\odot}})$. Marker size is proportional to $\mathcal{L}$, colour maps temperature \teff. The error bars in the lower right panel are only shown for our measurements as they are not available for the sample of \citet{simon-diaz_iacob_2017}. \\ \\ }
         \label{fig:theta_observed_all}
\end{figure*}

The macroturbulent velocities $\Theta$ of the full sample are shown as histograms and Kernel Density Estimates (KDE) in Fig.\ref{fig:theta_observed_hist}. The left panel shows the entire sample. The middle panel shows two subsamples separated by metallicity (dashed light grey for the LMC and dark grey for the Galaxy). The right panel shows two subsamples separated by the method of measurement (dashed light grey for the method we applied and dark grey for the one used in \citet{simon-diaz_iacob_2017}), which we treat separately to analyse potential systematic shifts due to methodological differences. One can see three peaks in the distributions around 20 \kms, 40 \kms, and 60--80 \kms. Their meaning becomes clear when comparing $\Theta$ to the other observables such as temperature and luminosity in Fig.\ref{fig:theta_observed_all}, where colour shows temperature, symbol size -- spectroscopic luminosity, and the symbol itself (circles and triangles) shows whether the measurement originates from us or is taken from \citet{simon-diaz_iacob_2017}, respectively. The first and the last peaks of the histograms are only visible in the Galactic subsample and belong to the main-sequence and evolved populations, respectively. This is seen in Fig.\ref{fig:theta_observed_all}, with the bulk of low-$\Theta$ objects corresponding to low-luminosity and high-\logg\ at all temperatures. The highest Galactic peak of 60--80 \kms\ corresponds to the evolved population with low-\logg\ and high luminosity at all temperatures.  The middle 40 \kms\ peak originates clearly from the LMC subsample and is analogous to the peak of 60--80 \kms\ shifted to lower values due to the difference in the metallicity of the subsamples. There is no second low peak in the LMC subsample since all the observed targets from the LMC fell in the supergiant regime (see left panel in Fig.\ref{fig:HRD1}).  The peak at 80 \kms, while seen in the KDE as a separate feature, belongs to the same group of galactic SG as can be seen in 
Fig. \ref{fig:theta_observed_all}.

Another notable feature is visible in the KDE in the right panel of Fig.\ref{fig:theta_observed_hist}. A small shift of the low-$\Theta$ peak occurs between our and \citet{simon-diaz_iacob_2017} measurements of the Galactic main-sequence stars. The same difference may be noticed in the lower right panel of Fig.\ref{fig:theta_observed_all} (showing a pairwise plot of $\Theta$ and spectroscopic luminosity) as a slight kink present around 2-2.5 in $\ log (\mathcal{L}/\mathcal{L}_{\odot})$. It is unclear whether the shift is caused by methodological differences or is due to differences in the parameter region covered by our samples. Our Galactic sample covers masses from 2.5 to 5 $M_{\odot}$, while the sample of \citet{simon-diaz_iacob_2017} starts at 4 $M_{\odot}$, hence the overlap is limited to a few stars only.  

The upper left panel of Fig.\ref{fig:theta_observed_all} shows relations between the projected rotational velocity \vsini~ and macroturbulent velocity $\Theta$. One can see qualitatively three populations showing a linear growth of  $\Theta$ with \vsini~, as noticed by \citet{simon-diaz_iacob_2017}, with different slopes for each population. The clear linear trends are likely caused by the ability of the FT method to separate rotational and macroturbulent broadenings, which strongly depends on whether they operate in different frequency ranges of the FT or not. The limitations of the FT approach for measuring rotational velocities are extensively discussed by \citet{aerts_use_2014} and \citet{Serebriakova2023}, thus we restrain from the detailed discussion of this matter. We note though that the FT method relies on several assumptions. First, it assumes that the total broadening may be represented by a convolution of the rotational profile with all other mechanisms as independent functions. Second, the rotational profile itself is thought to be a simplified function with a fixed linear limb darkening coefficient, which allows for the FT of the rotational profile of a given \vsini~ to be characterised by a single parameter -- the position of the first zero. Moreover, the position of the first zero in the FT should ideally not overlap features caused by other broadening functions, which implies rotation dominating over all other mechanisms. Lastly, the analysed line profiles are assumed to be symmetric. In reality, these assumptions are rarely met and the FT of the line profiles shows several zeros that are smeared out and are thus difficult to interpret. This causes, in our case, high degeneracy between $\Theta$ and \vsini, and leads to the large errors for $\Theta$ in the lower right panel of Fig.\ref{fig:theta_observed_all}. 

In all pairwise plots of Fig.\ref{fig:theta_observed_all}, the macroturbulent velocities $\Theta$ show the same trends of increasing with temperature and luminosity and decreasing with surface gravity. The $\Theta$ of Galactic stars are systematically higher than those of the LMC stars, although it is not clear whether the difference is caused by the metallicity and not the bias of the LMC sample that is overpopulated with evolved stars. 

\section{MESA models}\label{sec:mesa}

A grid of stellar structure and evolution models was computed with the {\sc mesa} code (\citet{Paxton2011}, \citet{Paxton2013}, \citet{Paxton2015}, \citet{Paxton2018}, \citet{Paxton2019}, \citet{Jermyn2023}) of version r15140. The chosen input physics is taken from \citet{Michielsen2023}, where a detailed description is provided. For convection, we use the Mixing Length Theory (MLT) with a mixing length parameter $\alpha_{\rm MLT}$=2.0. The Ledoux criterion for convection was adopted, with no semi-convection included. 
\citet{Michielsen2023} includes extra core-boundary mixing, implemented as both step-function increase of the core size by a factor $\alpha_{\rm ov}$ (often referred to as step overshooting parameter),  and by the exponentially decreasing mixing characterised by $f_{\rm ov}$ parameter. We set these parameters to 0.05 and 0.005, respectively, to keep the extra mixing minimal.

The grid was computed for the full mass range in our sample and uniform coverage in the HRD (2, 2.5, 3, 4, 5, 7, 9, 12, 15, 20, 25, 32, 40, 60, and 80 $M_{\odot}$). The evolution was stopped at log\teff = 3.8 for masses up to 9$M_{\odot}$ and log\teff = 4.0  for higher masses, which is dictated by the span of our observed parameter ranges. 

\begin{figure*}[!htbp]
   \begin{centering}
            {\includegraphics[clip,width=540pt,trim={0.0cm 0.0cm 0.0cm 0.0cm}]{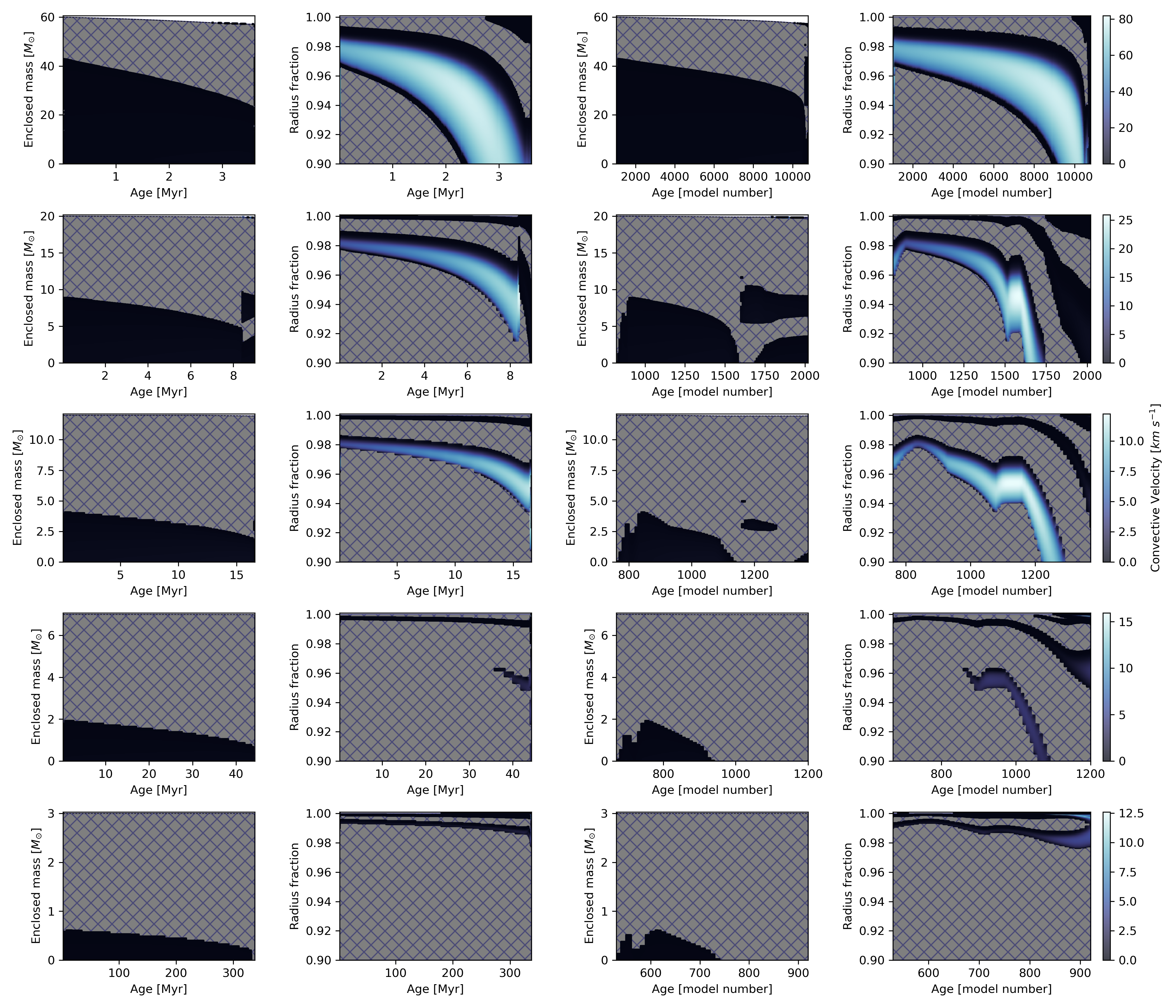}}
            
      \caption{ Kippenhahn diagrams for 60, 20, 12, 7, and 3 $M_{\odot}$ (rows) stellar models at metallicity $Z$=0.014. Odd columns -- full radius (in mass units), even columns -- zoom into the outer 10\% of radius (in radius units). Hatched areas denote radiative zones. The Ledoux-unstable convective zones are colour-mapped based on convective velocity. Two left columns are plotted with age in Myr, while two right columns are plotted with age in the model number, which allows a better view of the short-lived post-MS stages.}
         \label{fig:kippenhahn}
   \end{centering}
\end{figure*}

Fig.\ref{fig:kippenhahn} shows Kippenhahn diagrams for initial masses 60, 20, 12, 7, and 3 $M_{\odot}$ with $Z$=0.014 (while $Z$=0.006 are available in the Appendix \ref{app:kippenhahn}). The radiative transfer zones are hatched, and Ledoux-unstable convective zones are colour-mapped based on convective velocity. The odd panels show the upper 10\% of the stellar radius, where one can see the development of subsurface convective layers with age and initial mass, as well as ranges of the convective velocity. The latter is near-zero for lower masses and only reaches some 10~\kms~ late in the post-MS stage, while high-mass stars have up to 80 ~\kms~ already at the early MS.  

These models are shown in more detail in Figs. \ref{fig:z014_opacity_convvel_vs_temperature} and \ref{fig:z014_opacity_convvel_vs_radius_zoom} of Appendix \ref{app:mesaprofiles}, where full internal profiles are plotted for three snapshots at three evolutionary stages: mid-MS, TAMS, and HG. While the former plot shows the convective velocity and opacity versus temperature, the latter plot shows the same quantities as a function of the stellar radius. In Fig. \ref{fig:z014_opacity_convvel_vs_temperature}, one can clearly see the iron and helium opacity bumps depending on the mass and evolutionary stage, triggering the formation of a convective zone. The iron opacity bump at T $\sim 2\times10^5$\,K is responsible for the high convective velocities in the high mass regime (7 $M_{\odot}$ and higher). The \ion{He}{ii} opacity bump at T $\sim 4\times10^4$\,K causes convective velocities to occur at lower masses. This opacity bump also becomes significant at the late stages of evolution for all mass regimes in the expanded envelope, where it forms the second peak in the convective velocities with a much smaller amplitude than the velocities due to the iron peak. Despite the absolute values of the FeCZ convective velocities being large in some cases, we show that they are subsonic by plotting velocities as a fraction of the sound speed in Figs.\ref{fig:prof_z014_csound_convvel_vs_radius} and \ref{fig:prof_z006_csound_convvel_vs_radius}.

\section{Comparing convective velocity and macroturbulent velocity }\label{sec:conv_vs_vmacro}

\begin{figure*}
   \sidecaption
            {\includegraphics[clip,width=12cm,trim={1.0cm 0.5cm 1.0cm 1.0cm}]{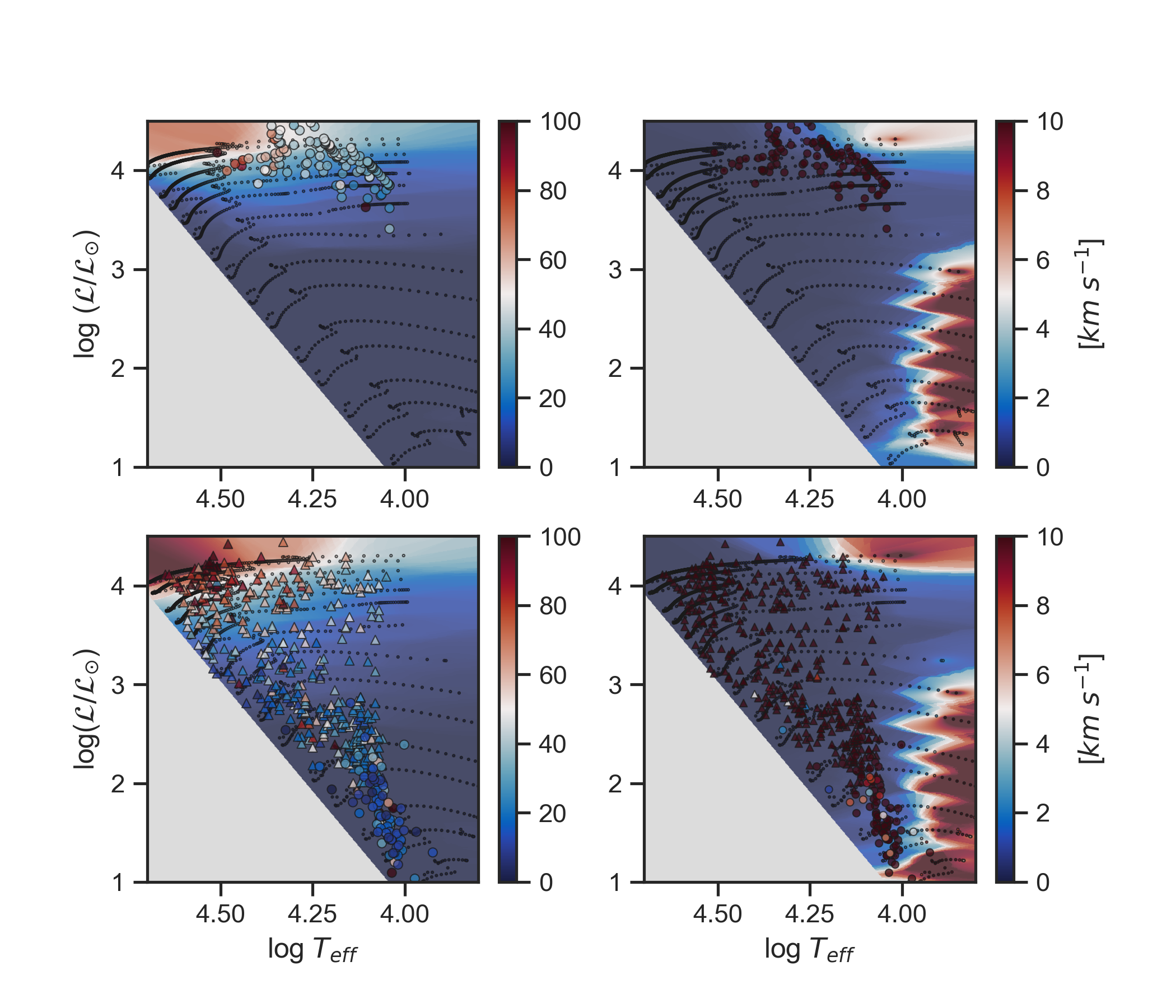}}
      \caption{ Convective velocities in the HRD interpolated from MESA models for continuous representation. Top to bottom: metallicity $Z$=0.006 and 0.014. Left to right: convective velocities in the regions of iron and helium opacity bumps. Evolutionary tracks that were used for the interpolation are shown in each panel. Coloured circles show observed macroturbulent velocities measured in this work, triangles those measured in \citet{Simon-Diaz2017}. \\ \\    }
         \label{fig:HR_vmac_FeHe}
\end{figure*}

\begin{figure*}
   \begin{centering}
            {\includegraphics[clip,width=500pt,trim={1.0cm 1.0cm 1.0cm 1.0cm}]{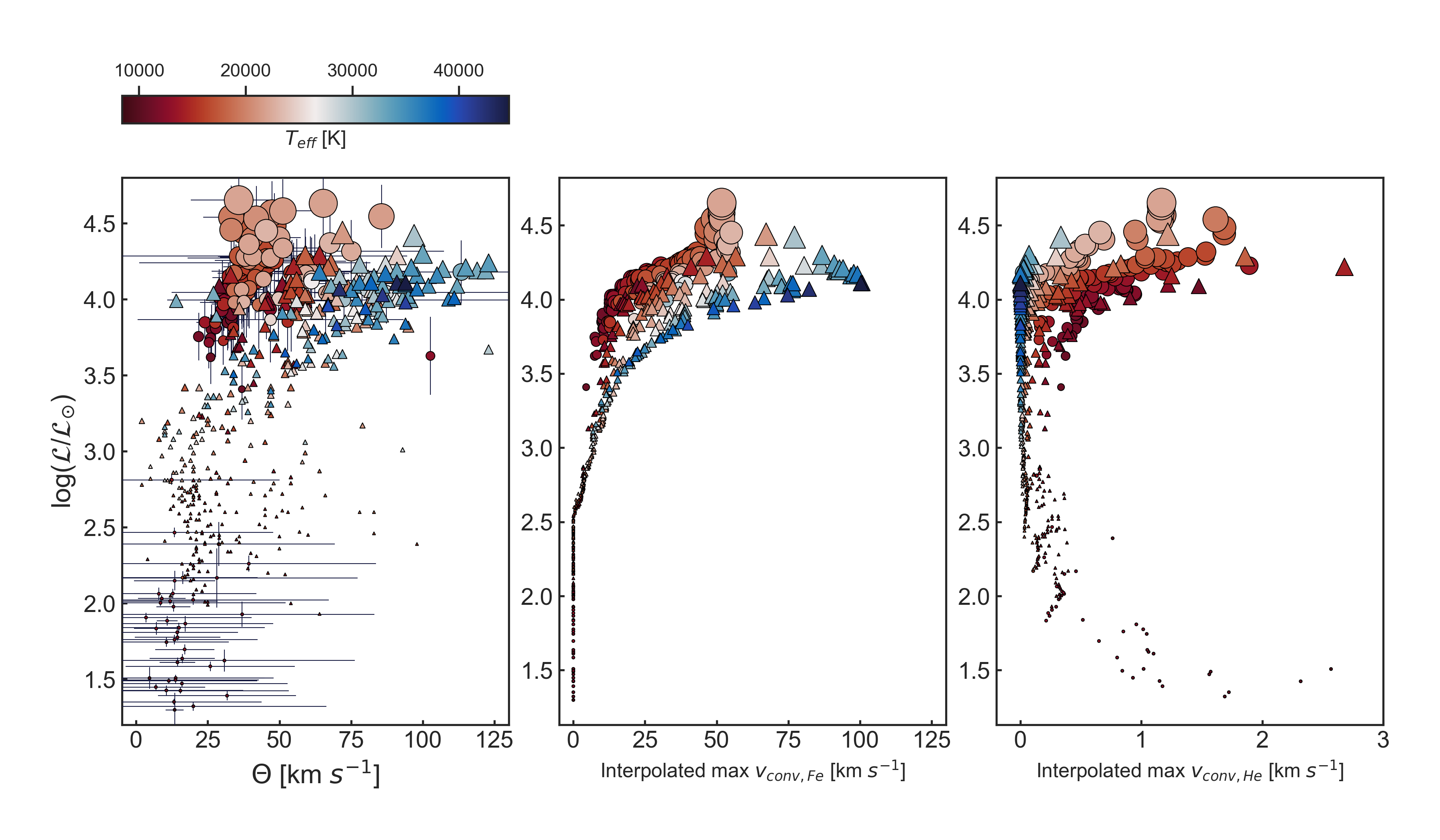}}
            
      \caption{ Dependence of the macroturbulent and convective velocities on spectroscopic luminosity of the star. From left to right: observed macroturbulent velocities (left) and predicted with MESA convective velocities for the iron (middle) and helium (right) opacity bumps. Convective velocities in the middle and right panels have been interpolated on the observed positions in the HRD from Fig.\ref{fig:HR_vmac_FeHe}. Symbol size and shape are the same as in Fig. \ref{fig:theta_observed_all}. }
         \label{fig:obs_vs_predict_vmac_FeHe}
   \end{centering}
\end{figure*}

Fig.\ref{fig:HR_vmac_FeHe} shows the distribution of the convective velocities as predicted by MESA models in the HRD, separately for the Fe and He opacity bumps. Measurements of the macroturbulent velocity are presented with the coloured symbols (see figure caption for details).  It is well visible that the convective velocities in the Fe opacity bump have very similar trends as the observed macroturbulent velocities in the high mass regimes ($M>10M_{\odot}$ or $\log (\mathcal{L} / \mathcal{L_{\odot}})>2.5$) with the same ranges of variation. For the lower masses, the observed macroturbulent velocities have much larger scatter and amplitudes than the convective velocities in both the Fe and He opacity bumps. The same is seen in Fig.\ref{fig:obs_vs_predict_vmac_FeHe} where the same macroturbulent and interpolated convective velocities are plotted along the x-axis vs spectroscopic luminosity 
$\log (\mathcal{L} / \mathcal{L_{\odot}})>2.5$, while colour maps the effective temperature. The He-bump velocity is too small to explain any observed macroturbulent velocities that are two orders of magnitudes larger. The Fe-bump only appears for stars with $\log (\mathcal{L} / \mathcal{L_{\odot}})>2.5$ or $M>10M_{\odot}$. Therefore, we hypothesise that the Fe-bump and macroturbulent velocities are connected in the high-mass regime while there is a lack of such relation for lower stellar masses.

Before embarking upon a comparison between the macroturbulence and the convective velocities, we point out that $\Theta$ measured from spectral lines cannot be directly fitted by the convective velocities in the subsurface convection layers of 1D models, since those layers are located much deeper in the stellar interior than the line formation region in the photosphere. This is a limitation of 1D evolution models based on mixing-length theory. \citet{schultz_stochastic_2022, schultz_turbulence-supported_2023} showed that subsurface velocities connected to turbulence do reach the photosphere in their 3D hydrodynamical simulations but these only cover dynamical time scales and cannot be used for evolutionary studies. Moreover, 2D unified box-in-a-star simulations by \citet{Debnath2024} reveal a 
highly structured and turbulent subsurface atmosphere causing 
turbulent velocities of order 30 to 100\,km\,s$^{-1}$ 
in the surface layers of galactic O stars. Their simulations 
give rise to turbulent gas interacting with a line-driven wind 
above the photosphere  and result in an outflow of 
gas rather than the parcels turning over and returning back into 
the stellar envelope. This interplay between the subsurface turbulent pressure and the stellar wind implies variations in the flux and creates turbulent velocities in the line-forming regions that scale linearly with the convective velocity obtained from MLT (Moens et al., in prep.). For this reason, it is a good strategy to analyse the possible connection between $\Theta$ and the convective velocities from 1D models based on MLT. We thus study their trends from a statistical point of view for a large sample of stars.

\subsection{Multivariate linear regression }\label{sec:conv_vs_vmacro_multivar}

As the first step of this analysis, we follow the statistical approach of \citet{Aerts2023} and \citet{Serebriakova2023} and employ multivariate linear regression. Multivariate linear regression is a statistical technique used to model the relationship between two or more predictors (independent variables) and a single predicted (dependent) variable. It extends the concept of simple univariate linear regression, which deals with one predictor and one predicted variable, to accommodate multiple predictors, thereby allowing for a more comprehensive analysis of complex data sets. Multivariate linear regression can be employed to investigate the relationships between stellar parameters and show how various factors simultaneously influence a variable of interest, which is, in our case, macroturbulent velocity $\Theta$. We investigate trends in the observed $\Theta$ quantitatively from multivariate linear regression models in the following form: 

\begin{equation}
    V = c_0 + \sum_{i=0}^{N} c_i  P_i ,
\end{equation}
where $V$ is a dependent variable (in our case, macroturbulent velocity $\Theta$), $P_i$ -- predictors (other observables or model parameters), and  $c_i$ -- coefficients of linear regression for a corresponding $i$-th predictor.

In \citet{Serebriakova2023}, based on adjusted $R^2$-statistics, 96.4\% of variability in the observed macroturbulent velocities of 126 LMC stars were found to be explained by only two predictors: \teff\ and \vsini, while the set of observables (\teff, \logg, $\log (\mathcal{L}/\mathcal{L}_{\odot})$, and \vsini) was tested and the rest observables were found insignificant for predicting $\Theta$. Due to the absence of any analogue of \vsini\ in our non-rotating {\sc mesa} models, we need to repeat the test of different combinations of predictors without \vsini\ to find a set of significant observables that are available in models. In absence of \vsini, we found both  \logg\ and $\log (\mathcal{L}/\mathcal{L}_{\odot})$, in addition to log\teff, to be significant, although $\log (\mathcal{L}/\mathcal{L}_{\odot})$ is computed from $\mathcal{L} = T_{\text{eff}}^4 / g$. Therefore, we further test all three parameters, $\log (\mathcal{L}/\mathcal{L}_{\odot})$, log\teff, and \logg, as predictors on the observed data. The results of the test are presented in Table \ref{tab:multivar_obs} as p-values and coefficients $c_i$ per each predictor, and adjusted $R^2$ per tested dataset. 

\begin{table*}
\centering
\caption{Results of the multivariate linear regression.  }
\begin{tabular}{lcccccc}
\toprule
Dataset & $\log (\mathcal{L}/\mathcal{L_{\odot}})$ & \teff & \logg & \text{Const} & \textbf{Adj. $R^2$} \\
\midrule

\multirow{2}{*}{Obs Gal, all $ \log (\mathcal{L}/\mathcal{L}_{\odot})$}  & \multirow{2}{*}{0.000} & \multirow{2}{*}{0.000} & \multirow{2}{*}{0.000} & \multirow{2}{*}{0.000} & \multirow{2}{*}{\textbf{0.331}} \\ & 25.72 & -15.33 & 12.00 & -9.33 & \\

\multirow{2}{*}{Obs Gal, $ \log (\mathcal{L}/\mathcal{L}_{\odot}) > 2.5$} & \multirow{2}{*}{0.000} & \multirow{2}{*}{0.000} & \multirow{2}{*}{0.000} & \multirow{2}{*}{0.000} & \multirow{2}{*}{\textbf{0.569}} \\ & 49.73 & -37.11 & 18.11 & -20.37 & \\

\multirow{2}{*}{Obs Gal, $ \log (\mathcal{L}/\mathcal{L}_{\odot}) > 3.0$}  & \multirow{2}{*}{0.000} & \multirow{2}{*}{0.000} & \multirow{2}{*}{0.000} & \multirow{2}{*}{0.000} & \multirow{2}{*}{\textbf{0.602}} \\ & 79.08 & -68.56 & 31.23 & -36.22 & \\

\multirow{2}{*}{Obs Gal, $ \log (\mathcal{L}/\mathcal{L}_{\odot})  < 3.0$} & \multirow{2}{*}{0.310} & \multirow{2}{*}{0.244} & \multirow{2}{*}{0.899} & \multirow{2}{*}{0.322} & \multirow{2}{*}{\textbf{0.001}} \\ & -5.52 & 9.78 & -0.97 & 4.30 & \\

\multirow{2}{*}{Obs Gal, $ \log (\mathcal{L}/\mathcal{L}_{\odot})  < 2.5$}  & \multirow{2}{*}{0.568} & \multirow{2}{*}{0.524} & \multirow{2}{*}{0.263} & \multirow{2}{*}{0.477} & \multirow{2}{*}{\textbf{0.002}} \\ & 7.30 & -11.98 & 18.78 & -6.97 & \\ \hline

\multirow{2}{*}{Obs LMC, all $\log (\mathcal{L}/\mathcal{L}_{\odot})$} & \multirow{2}{*}{0.002} & \multirow{2}{*}{0.001} & \multirow{2}{*}{0.000} & \multirow{2}{*}{0.000} & \multirow{2}{*}{\textbf{0.420}} \\ & 20.67 & -21.48 & 28.72 & -12.75 & \\

\bottomrule
\end{tabular}
\tablefoot{
The quantities $\log (\mathcal{L}/\mathcal{L_{\odot}})$, \teff, and \logg~ were used as predictors and observed macroturbulent velocities as the dependent variable. The rows correspond to different subsets of the data separated by a range of spectroscopic luminosities and by metallicity. $P>|t|$ and $c_i$ values for each predictor are given in every cell as top and bottom numbers, respectively. The adjusted $R^2$ are given in the last column.
}
\label{tab:multivar_obs}
\end{table*}

We repeated the exercise for various subsamples with cut-off by $\log (\mathcal{L}/\mathcal{L}_{\odot})$ in order to look at behaviour of the measured microturbulent velocities in different luminosity regimes. These regimes are evident in Fig.\ref{fig:obs_vs_predict_vmac_FeHe} where a turn of $\Theta$ is seen towards higher values above $\log (\mathcal{L}/\mathcal{L}_{\odot})$ = 2.5. For the LMC sample, we only use the full sample as all stars in that sample already have $\log (\mathcal{L}/\mathcal{L}_{\odot})$ > 3. In Table \ref{tab:multivar_obs}, one can see how the fraction of explained variance (reflected in adjusted $R^2$) increases for subsamples with higher $\log (\mathcal{L}/\mathcal{L}_{\odot})$, and reaches 60\%. Moreover, the samples that only include stars with $\log (\mathcal{L}/\mathcal{L}_{\odot})$ < 3 show large p-values and near-zero $R^2$, suggesting the predictors fail to explain variability in $\Theta$ for low-$\log (\mathcal{L}/\mathcal{L}_{\odot})$ subsample.

Furthermore, we can compare the roles of each predictor in explaining observed $\Theta$  with the roles of the same predictors in explaining model convective velocities $v_{conv}$ by looking at their coefficients. We repeated the same multivariate linear regression for the model sample, using each computed internal profile as a point of a sample with model $v_{conv}$ as the dependent variable. First, we repeated tests of sets of predictors for their predicting power (which are not limited by a modest set of observables that we have to stick to with observed data). While it appeared possible to describe up to 96\% of variability using model predictors such as age and mass (as expected), a subset of observable parameters had to be selected for qualitative comparison with observations. Same $\log (\mathcal{L}/\mathcal{L}_{\odot})$, log\teff, and \logg\ were found to all be significant and explain up to 79\% of variability in the model subsets. The results of multivariate linear regression of the model data are presented in the Appendix \ref{app:tables} in Tables \ref{tab:multivar_max_all} and \ref{tab:multivar_integr_all} as the model data provided more possible subsets. The subsets include both metallicities with full coverage of the parameter space, convective velocity in  both iron ($v_{\rm conv, Fe}$) and helium  ($v_{\rm conv, He}$) convective layers, as well as maximum ($v_{\rm conv, x}^{\rm max}$) or integrated velocity ($v_{\rm conv, x}^{\rm integr}$) chosen as dependent variable. 

The higher adjusted $R^2$ values in the model data compared to observations (for example of $v_{\rm conv, Fe}^{\rm max}$ subsets with $\log (\mathcal{L}/\mathcal{L}_{\odot})$ > 3, $R^2$ is 0.794 against 0.602 for the Solar metallicity, or 0.722 against 0.420 for the LMC metallicity) suggest  additional complexities (rotation) or observational uncertainties (in both the position in HRD and measured $\Theta$) in the observed datasets.  The overall consistency in coefficient trends between the model and observed datasets for each predictor suggests that the fundamental physical relationships between the predictors and the $v_{\rm conv, Fe}^{\rm max}$ and $\Theta$ are similar.  The similar coefficient trends across different luminosities and metallicities also indicate that the predictors $\log (\mathcal{L}/\mathcal{L}{\odot})$, \teff, and \logg\ play comparable roles in influencing both macroturbulent and convective velocities, which was seen in Fig. \ref{fig:HR_vmac_FeHe} and \ref{fig:obs_vs_predict_vmac_FeHe}.  The largest discrepancies are found in low-$\log (\mathcal{L}/\mathcal{L}_{\odot})$, where model dataset have near-zero $v_{\rm conv, Fe}^{\rm max}$ or $v_{\rm conv, He}^{\rm max}$ with no scatter, while observations have a large scatter around still noticeable velocities with weak to no trends with other observables.   

The comparison of all tests shows that the observed scatter in macroturbulent velocities $\Theta$ for the low-mass regime ($\log (\mathcal{L}/\mathcal{L}_{\odot})$ < 3 or $M<12M_{\odot}$) can not be explained by same predictors that partially describe model convective velocities. For high-mass regime ($\log (\mathcal{L}/\mathcal{L}_{\odot})$ > 3 or $M>12M_{\odot}$) and Solar metallicity, 60.2\% of the observed $\Theta$ variability is explained by the same predictors and similar coefficients as Fe-bump convective velocities, pointing to the common origin of the two phenomena. Even though it is noticeably less than 79.4\% described in the model convective velocities, the observed sample has a scatter due to the measurement errors that smear out the trends. Moreover, the coefficients of all predictors show consistently similar trends with luminosity and metallicity, which illustrates that the predictors play comparable roles in influencing both macroturbulent and convective velocities in the observations and models, respectively.

\subsection{Principal component analysis }\label{sec:conv_vs_vmacro_pca}

Principal Component Analysis (PCA) is a statistical technique widely utilised in the context of multivariate data analysis to investigate the underlying structure and trends within high-dimensional datasets. PCA operates by transforming the original n-dimensional data into a new coordinate system where the axes (principal components) are ordered by the amount of variance they capture from the data. The first principal component captures the maximum variance, the second captures the second most variance, and so on. This transformation is achieved through an eigenvalue decomposition of the covariance matrix of the data, yielding eigenvalues and eigenvectors that define the principal components.

In the context of comparing two n-dimensional point clouds, PCA can be applied separately to each dataset to derive their respective principal components. By examining the principal components, we can compare the direction and magnitude of the variance captured in each dataset. Specifically, if the principal components of both clouds align similarly (i.e., they have similar eigenvectors and eigenvalues), it suggests that the two datasets share similar trends. Conversely, significant discrepancies in the principal components would indicate divergent trends between the two point clouds.

In  order to compare the trends in $\Theta$ and $v^{\rm max}_{\rm conv, Fe}$, we independently apply PCA to four clouds of points: observed LMC, model LMC, observed Galactic, and model Galactic datasets. In each case, we used same variables, with the difference that $\Theta$ and $v^{\rm max}_{\rm conv, Fe}$ were only used in the observed and model data, respectively, so the full list of dimensions per each point in the cloud consisted of \teff, \logg, $\log (\mathcal{L}/\mathcal{L_{\odot}})$, and either $\Theta$ or $v^{\rm max}_{\rm conv, Fe}$. We included the first two components only since they provided cumulative explained variances of 89 to 95 \% in each of the four datasets, as with the ratio per component shown in Table \ref{tab:variance}.

\begin{table*}
\centering
\caption{Explained variance ratios and cumulative explained variance}
\begin{tabular}{lcccccccc}
\toprule
  & \multicolumn{2}{c}{\textbf{LMC Obs}} & \multicolumn{2}{c}{\textbf{LMC Model}} & \multicolumn{2}{c}{\textbf{Gal Obs}} & \multicolumn{2}{c}{\textbf{Gal Model}} \\
 & PC1 & PC2 & PC1 & PC2 & PC1 & PC2 & PC1 & PC2 \\
\midrule
Explained Variance Ratio & 0.6487 & 0.2429 & 0.5680 & 0.3857 & 0.5507 & 0.3756 & 0.6063 & 0.3472 \\
Cumulative Explained Variance & 0.6487 & \textbf{0.8916} & 0.5680 & \textbf{0.9536} & 0.5507 & \textbf{0.9263} & 0.6063 & \textbf{0.9535} \\
\bottomrule
\end{tabular}
\label{tab:variance}
\end{table*}

Each principal component may be represented as a linear combination of the initial set of parameters (\teff, \logg, $\log (\mathcal{L}/\mathcal{L_{\odot}})$, and either $\Theta$ or $v^{\rm max}_{\rm conv, Fe}$). The coefficients for each parameter in each principal component for each dataset are provided in Table \ref{tab:loadings_and_correlations}. Comparison of these coefficients for four datasets allows us to draw the following conclusions. The primary component (PC1) shows consistent contributions from all parameters across all datasets, indicating that the primary trend is similar in both observations and models. The coefficients for $v^{\rm max}_{\rm conv, Fe}$ and $\Theta$ are comparable, suggesting these parameters are interchangeable in capturing the primary trend. However, the secondary component (PC2) exhibits significant differences in both the sign and magnitude of the coefficients for most parameters. These discrepancies may be attributed to specifics of the observed sample, such as uncertainties, presence of extra parameter \vsini, that is absent in the models, and limited sample size and parameter span.

\begin{table*}
\centering
\caption{Principal component loadings and correlations}
\begin{tabular}{lcccccccc}
\toprule
\textbf{Variable} & \multicolumn{2}{c}{\textbf{LMC Obs}} & \multicolumn{2}{c}{\textbf{LMC Model}} & \multicolumn{2}{c}{\textbf{Gal Obs}} & \multicolumn{2}{c}{\textbf{Gal Model}} \\
 & PC1 & PC2 & PC1 & PC2 & PC1 & PC2 & PC1 & PC2 \\
\midrule
\multicolumn{9}{l}{\textbf{Loadings}} \\
$T_{\text{eff}}$ & 0.6052 & 0.0236 & 0.5786 & -0.3787 & 0.5060 & -0.5168 & 0.5835 & -0.3338 \\
$\log g$ & 0.5505 & -0.3848 & 0.4407 & -0.5961 & -0.0594 & -0.8095 & 0.3568 & -0.7008 \\
$\log (\mathcal{L}/\mathcal{L_{\odot}})$ & 0.2604 & 0.9191 & 0.4614 & 0.5394 & 0.6103 & 0.2653 & 0.4794 & 0.5217 \\
$\Theta$ & 0.5127 & -0.0814 &  &  & 0.6065 & 0.0849 &  &   \\
$v_{\text{conv, Fe}}^{\text{max}}$ &  &  & 0.5080 & 0.4586 &  &  & 0.5500 & 0.3540 \\
\midrule
\multicolumn{9}{l}{\textbf{Correlations}} \\
$T_{\text{eff}}$ & 0.9749 & 0.0233 & 0.8721 & -0.4703 & 0.7510 & -0.6335 & 0.9086 & -0.3934 \\
$\log g$ & 0.8868 & -0.3793 & 0.6643 & -0.7404 & -0.0882 & -0.9922 & 0.5556 & -0.8259 \\
$\log (\mathcal{L}/\mathcal{L_{\odot}})$ & 0.4194 & 0.9059 & 0.6955 & 0.6700 & 0.9059 & 0.3252 & 0.7465 & 0.6149 \\
$\Theta$ & 0.8259 & -0.0802 &   &   & 0.9002 & 0.1041 &   &  \\
$v_{\text{conv, Fe}}^{\text{max}}$ &  &  & 0.7657 & 0.5695 &  &  & 0.8565 & 0.4172 \\
\bottomrule
\end{tabular}
\label{tab:loadings_and_correlations}
\end{table*}

\begin{figure*}
   \begin{centering}
            {\includegraphics[clip,width=240pt,trim={0.0cm 0.0cm 0.0cm 0.0cm}]{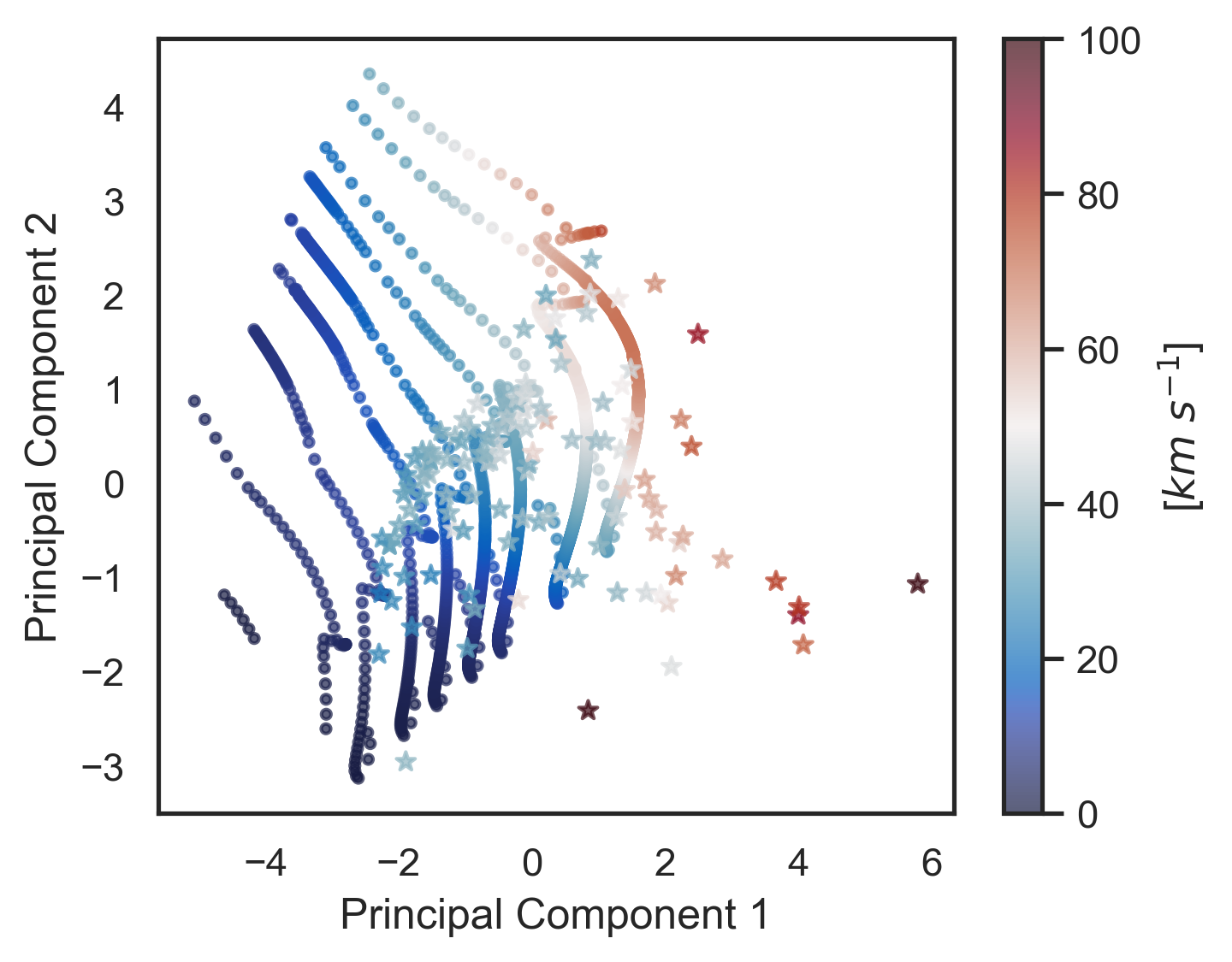}}
            {\includegraphics[clip,width=240pt,trim={0.0cm 0.0cm 0.0cm 0.0cm}]{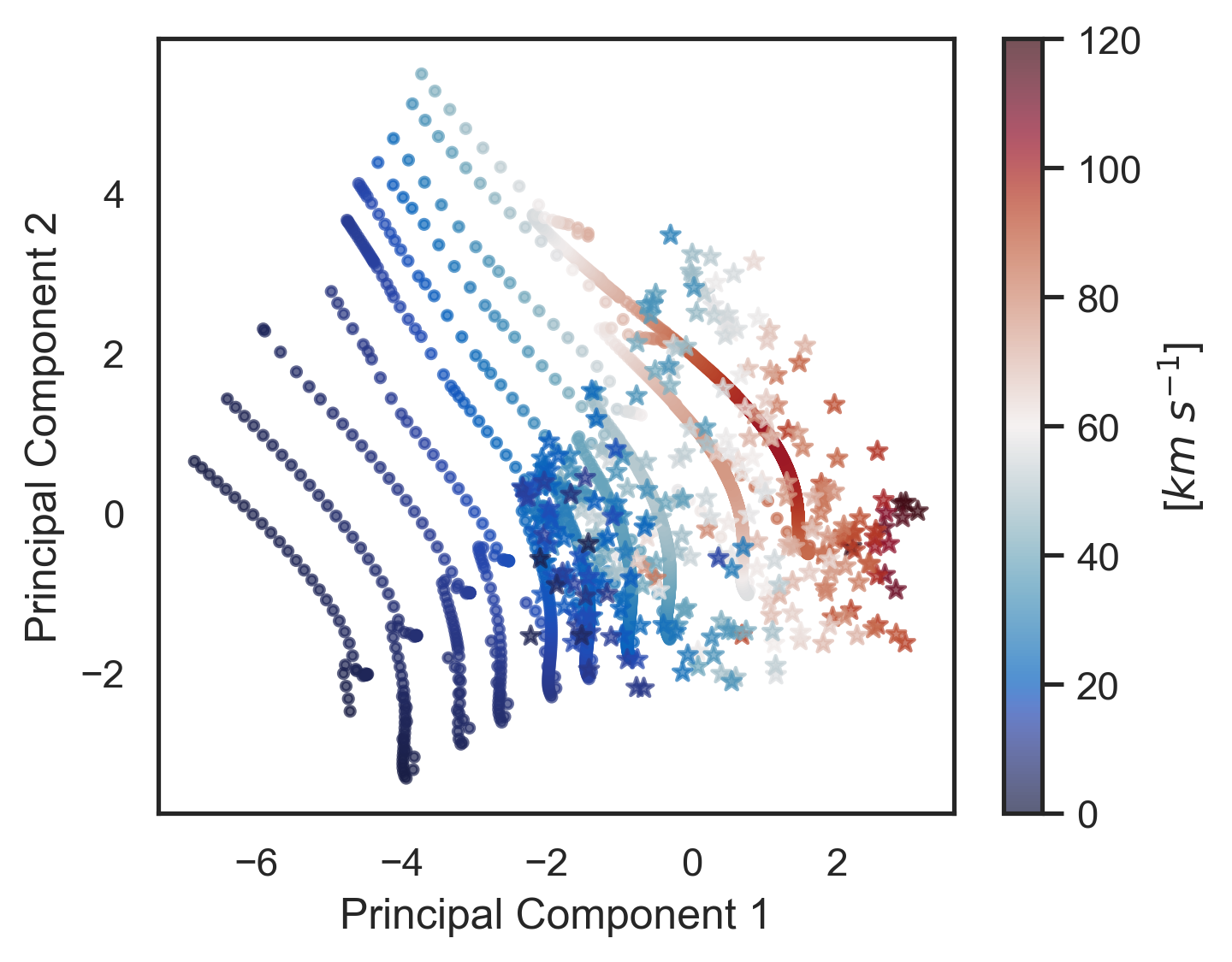}}
            
      \caption{ Datasets projection on the principal components space. Left panel -- models with $Z$=0.006 (circles) and observed LMC sample (stars), right panel -- models with $Z$=0.014 (circles) and observed Galactic sample (stars). Colour shows convective velocity for the model data and macroturbulent velocities for the observed data. All the four datasets were analysed independently. }
         \label{fig:pca}
   \end{centering}
\end{figure*}

The visual representation of all four datasets in PC1-PC2 space is shown in Fig.\ref{fig:pca}, where circles represent model data points, and stars represent observational data points. The colour indicates the value of $\Theta$ for observations and $v^{\rm max}_{\rm conv, Fe}$ for models. Although the four datasets were analysed independently, they align very similarly in the PC1-PC2 space. The alignment of points and matching colour gradients strongly indicate that the primary trend is consistent across all four datasets.
This suggests that the underlying structure and trends in the full 4-dimensional clouds are captured similarly by the first principal component across all datasets, supporting the hypothesis of $\Theta$ and $v^{\rm max}_{\rm conv, Fe}$ playing the same role in forming the trends. On the contrary, if we compare how the observed sample is misaligned from model tracks (compared to how the plot looks in the HRD instead of PC1-PC2), we can see obvious stretch and shrink of spans of the observed samples. Looking at the Galactic sample (second panel in Fig.\ref{fig:pca}), one can see the shift and shrink of the observed sample while matching colours. The shift indicates systematic differences in absolute values of $\Theta$ and $v^{\rm max}_{\rm conv, Fe}$ with the former being systematically higher. The shrink of the cloud to match lower-masses observations with higher-masses tracks indicates that the observed $\Theta$ in lower-mass regime is inconsistent with near-zero convective velocities of the models, while high-mass observed points are kept aligned with high-mass tracks. We interpret this as a support for the hypothesis of having two principally different regimes, 
with $\Theta$ in high- and low-mass regimes caused by different mechanisms.
The LMC sample, in contrast to the Galactic sample, seems to be stretched compared to model tracks and misaligned in colour, which we interpret to be caused by the small sample size in a very limited span of masses, while the Galactic sample covers the full range of parameters.


\section{On wave tunnelling through subsurface convective layers }\label{sec:igw}

\begin{figure*}
   \begin{centering}
            {\includegraphics[clip,width=490pt,trim={0.0cm 0.0cm 0.0cm 0.0cm}]{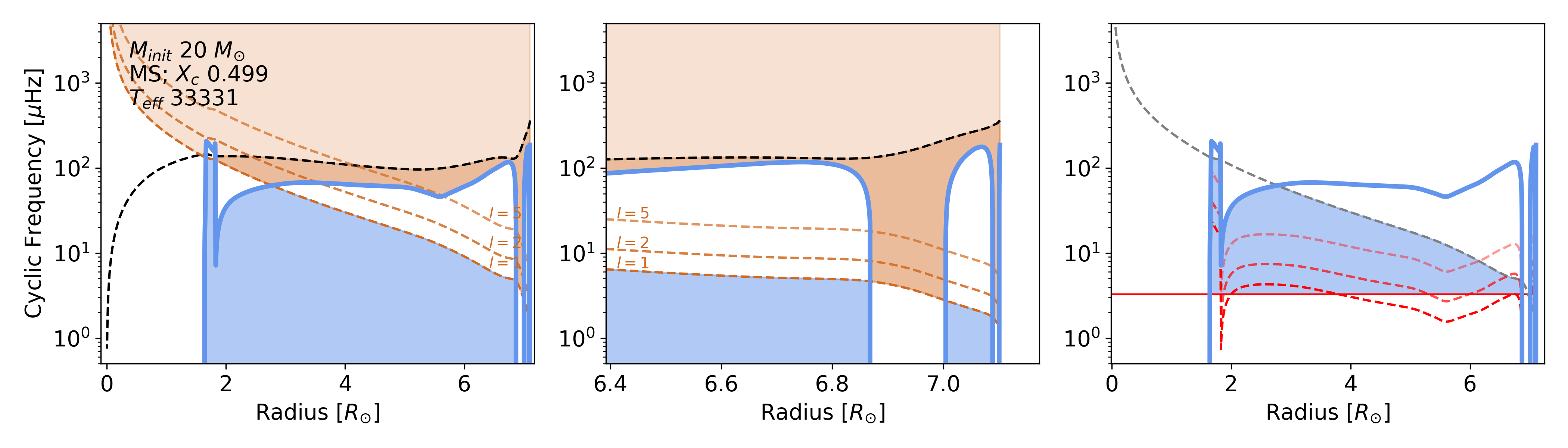}}
            
      \caption{ Propagation diagram for a 20 $M_\odot$ main-sequence star model. The x-axis represents the radius in solar radii, y-axis shows the cyclic frequency in microhertz. The left and right panels show the full radius, while the middle panel shows the external 10\% of the radius.  The blue line represents the Brunt-Väisälä frequency $N$, the brown dashed lines correspond to the Lamb frequencies $S_l$ for different spherical harmonic degrees $l$, and the black dashed line shows the profile of the acoustic cut-off frequency. The blue-shaded region shows the g-mode cavity, and the brown-shaded region -- p-mode cavity. The right panel shows profiles of tunnelling $\omega_{min}$ for $l$=1,2,5 in dashed red lines and one horizontal value of $\omega_{min}$ taken near the convective layer (see text for details). Profiles of $\omega_{min}$ collapse to an apparent vertical line in the He convection zone, and this particular zone is better resolved in zoomed version of the figure in Appendix D.
 }
         \label{fig:propagation_single}
   \end{centering}
\end{figure*}

Model internal profiles computed with {\sc mesa} allow us to look into the wave propagation by comparing the Brunt-Väisälä $N$ and Lamb $S_l$ frequencies across the stellar radius. The first two panels of Fig.\ref{fig:propagation_single} show an example of such a propagation diagram, i.e. internal profiles of $N$ and $S_l$ for a 20 $M_\odot$ main-sequence star model at the Solar metallicity. The blue line shows $N$, the dashed brown lines -- $S_l$ for $l$=1,2,5, and the black dashed line -- the acoustic cut-off frequency. The left panel shows the full radius, and the middle one -- zoom-in into the outer 10\% of the envelope. The blue-shaded area indicates regions where g-modes and IGW can propagate. This region is defined by frequencies $\omega$ less than both $N$ and $S_l$ and where $N^2$ is positive (in radiative layers, as negative corresponds to convective layers where g-modes cannot propagate). The brown shaded area shows regions where acoustic modes and waves can propagate, which is defined by frequencies $\omega$ greater than both the Brunt-Väisälä frequency and the Lamb frequency, but less than the acoustic cut-off frequency. Finally, the white regions where $N$ < $\omega$ < $S_l$ or $S_l$ < $\omega$ < $N$ represent evanescent zones, where waves exponentially decay. In this particular model, one can see present g-mode and p-mode cavities between the core boundary at 2 $R_\odot$ and iron-bump convective zone FeCZ at some 6.85 $R_\odot$, as well as some between the upper boundary of the FeCZ and the lower boundary of the HeCZ. However, a combination of the extensive FeCZ between 6.85 and 7 $R_\odot$ with rapidly decreasing $S_l$ and next HeCZ right below the surface does not leave any potential $\omega$ from g-modes or p-modes cavities that could reach the surface. The same propagation plots for models of different masses and evolutionary states are provided in Appendix \ref{app:propagation} (Figs. \ref{fig:prof_prop_simple_z014} and \ref{fig:prof_prop_simple_z014_zoom} for the Solar metallicity and Figs. \ref{fig:prof_prop_simple_z006} and \ref{fig:prof_prop_simple_z006_zoom} for the LMC metallicity).

As a useful addition to the propagation diagrams, we can roughly estimate the minimal wavelength a wave must have to tunnel through the subsurface convective layers before being decayed. The minimal requirement of the wavelength $\lambda$ being larger than the span of the region we want the waves to tunnel through may be expressed through wavenumber $k$ as: 

\begin{equation}
\lambda = \frac{2\pi}{|k|}.
\end{equation}

The frequency $\omega$ is connected to the wavenumber through the dissipation relation of IGW (\citet{asteroseismology}):

\begin{equation}
\omega^2 = \frac{k_h^2 N^2}{k_r^2 + k_h^2}.
\end{equation}

Here,  $k_h$ and $k_r$ are the radial and horizontal components of wavenumber $|k|^2 = k_r^2 + k_h^2$

The horizontal wavenumber $k_h$ can be expressed in terms of the spherical harmonic degree $l$ and radius $r$ as:

\begin{equation}
k_h = \sqrt{\frac{l(l+1)}{r^2}}.
\end{equation}

The condition $\lambda$ > $r_2-r_1$, where $r_1$ and $r_2$ are the positions of lower and upper boundaries of a convective layer, gives the minimal frequency allowing IGW to tunnel through the evanescent zone.

\begin{equation}
\omega_{min} = \frac{ (r_2-r_1) N(r) \sqrt{l(l+1)}}{2 \pi r}.
\end{equation}

The last equation shows that 'tunnelling' frequency follows the profile of the Brunt-Väisälä frequency scaled by relations of $r_2-r_1$ and radius $r$, for degrees $l$. The right panel of Fig.\ref{fig:propagation_single} shows internal profile $\omega_{min}$ for $l$=1,2,5 as red dashed lines, where the span of FeCZ was taken as $r_2-r_1$. A single minimum may be taken as a value of the $\omega_{min}$ profile for $l$=1 at a point before the convective zone and is shown in the same figure as a solid red horizontal line. This gives a minimal value for a frequency below which waves from below the convective layer can not tunnel through the evanescent zone. This may serve as an additional constraint for the g-mode cavity, extending the condition $\omega$ < $N$ < $S_l$ into $\omega_{min, l=1, r=r1^*}$ < $\omega$ < $N$ < $S_l$. Frequencies satisfying this condition occur in the blue-shaded regions in the right panel of Fig. \ref{fig:propagation_single}. Same modified propagation diagrams with 'tunnelling' $\omega_{min}$ for models of different masses and evolutionary states are provided in the Appendix \ref{app:propagation} (Figs. \ref{fig:prof_prop_wmin_z014} and \ref{fig:prof_prop_wmin_z006}  for the Solar and LMC metallicities, respectively). Note that $r_2-r_1$ was considered individually for each model based on which subsurface convective layer is the most extended for a given mass and age, i.e. the most restrictive on $\lambda$. Such a definition also allows us to introduce a flag for each model depending on the extent of the blue-shaded region between the core boundary and the lower boundary of the thickest convective layer. In other words, if a certain model does not allow any $\omega$ satisfying $\omega_{min, l=1, r=r1^*}$ < $\omega$ < $N$ < $S_l$ due to $r_2-r_1$ requiring $\omega_{min}$ > $N$ (or > $S_l$), we can flag such a model as forbidding IGWs to tunnel through to the surface. Note that such a crude definition of $\omega_{min}$ makes it rather conservative and may only be used to mark models that forbid tunnelling, but not to say that the rest of the models allow for it.

\begin{figure*}
   \begin{centering}
            {\includegraphics[clip,width=540pt,trim={1.0cm 0.0cm 1.0cm 1.0cm}]{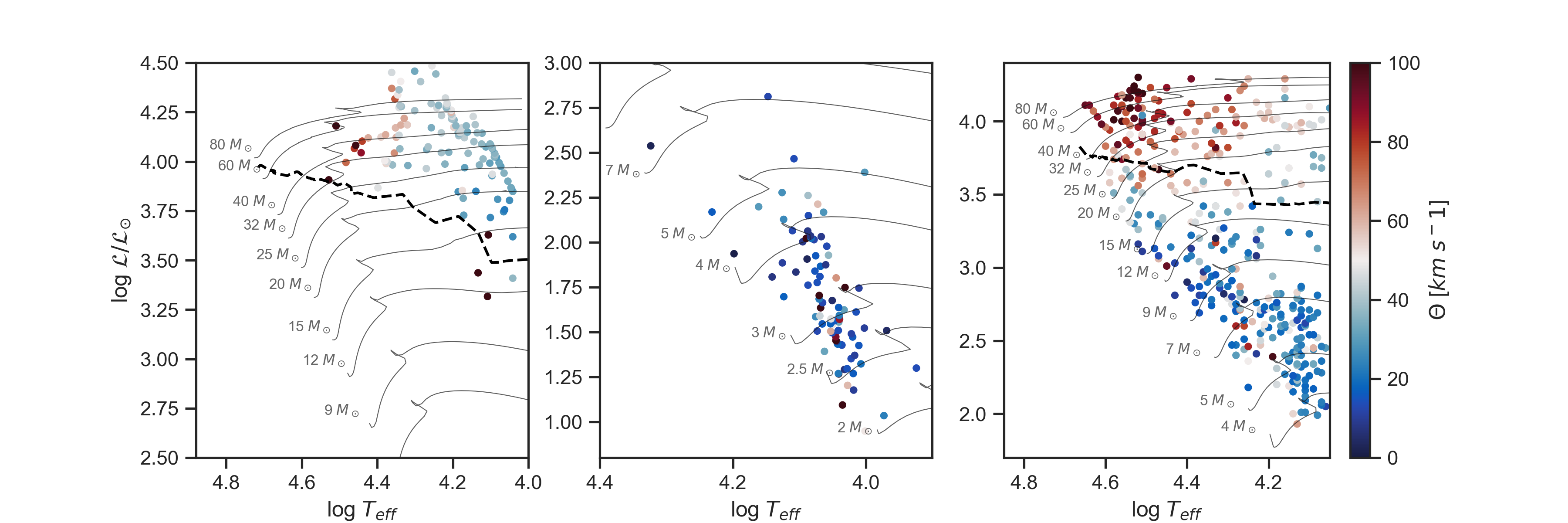}}
            
      \caption{ The observed sample with panels as in Fig.\ref{fig:HRD1}: left - for the LMC sample and metallicity, middle and right - for the Galactic samples and Solar metallicity. Colour shows observed macroturbulent velocities $\Theta$. The thick dashed black line separates the region of forbidden tunnelling (above the line) through subsurface convective layers (see text for details). }
         \label{fig:hrd_tunnelling}
   \end{centering}
\end{figure*}

\begin{figure*}
 \sidecaption
            {\includegraphics[clip,width=12cm,trim={0.8cm 1.0cm 1.5cm 0.5cm}]{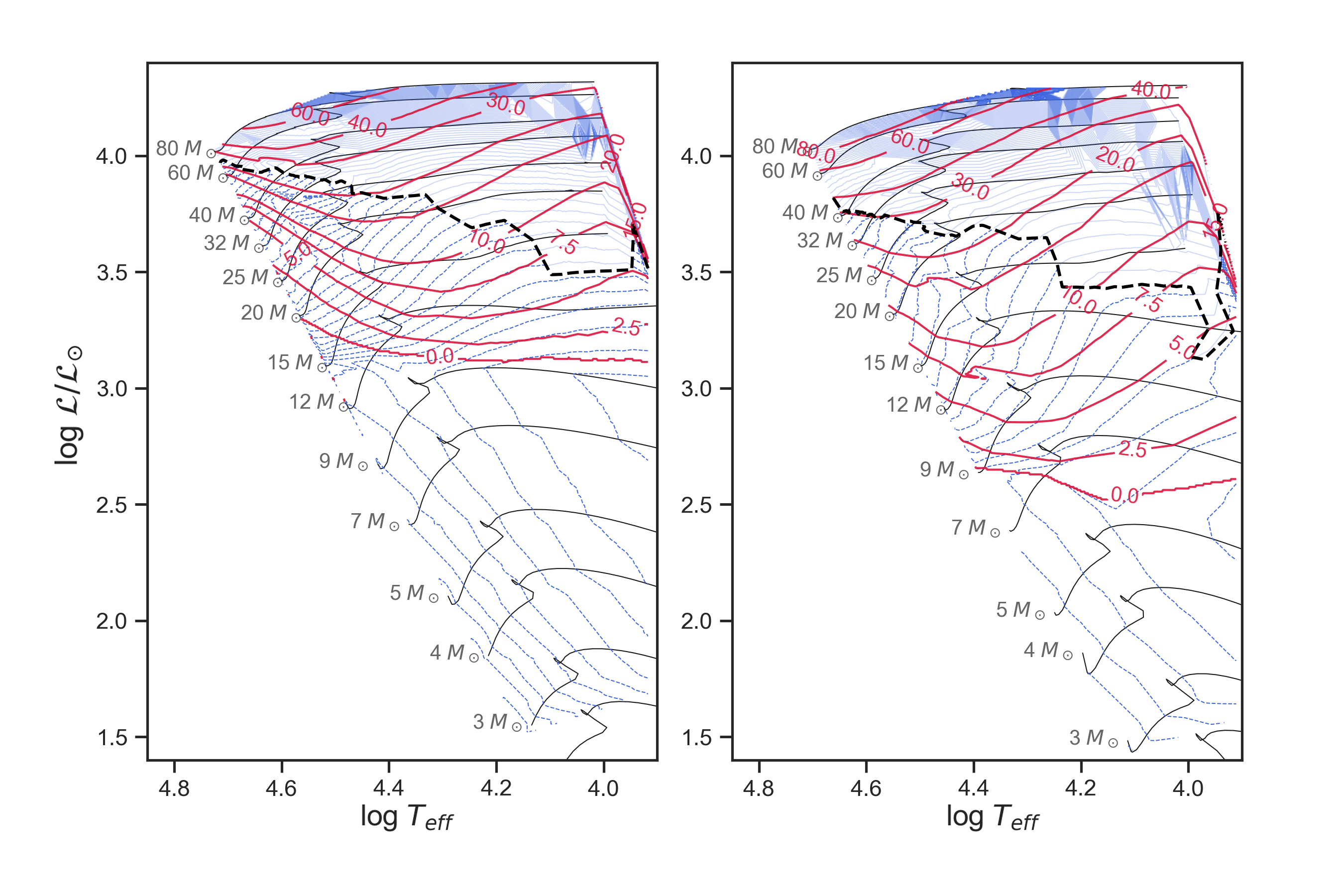}}
      \caption{ Level-plot of FeCZ maximal convective velocities in solid red lines and tunnelling frequency $\omega_{min}$ in dashed blue lines (see text for details). The thick dashed black line separates the region of forbidden tunnelling (above the line) through subsurface convective layers. \\ \\  }
         \label{fig:levels_tun_conv}
\end{figure*}

Based on the above definition of models with forbidden tunnelling, we may define corresponding regions in the HRD and compare them with the observed sample. Fig.\ref{fig:hrd_tunnelling} shows the full observed sample in the HRD with colour representing observed macroturbulent velocities $\Theta$ on top of evolutionary tracks. A thick black dashed line defines a boundary between models that forbid tunnelling of core-excited IGWs through subsurface convective layers (above the line) and the models not forbidding it (below the line). One can see that the vast majority of observed high-macroturbulence stars lie above the line, where tunnelling is not possible. This is a strong evidence against core-generated IGWs being responsible for high macroturbulent velocities observed in blue supergiants. This region occurs above 40 $M_\odot$ at ZAMS, 20 $M_\odot$ at TAMS, and 12 $M_\odot$ in the Hertzsprung gap for the Solar metallicity, and above 60 $M_\odot$ at ZAMS, 32 $M_\odot$ at TAMS, and 12 $M_\odot$ in the Hertzsprung gap for the LMC metallicity. in all these cases, the FeCZ is extended enough to block waves from the inside. These cases also have convective velocities up to 80 \kms, sufficiently large to describe the high observed macroturbulent velocities.

While the region in the lower HRD below the dashed line lacks truly high-macroturbulent stars (except several outliers), the observed microturbulent velocities almost never fall below 20 \kms. In the same region, models predict (near-)zero convective velocities (see Fig.\ref{fig:levels_tun_conv}), rendering the FeCZ unlikely to cause the observed macroturbulent broadening, yet allowing for core-generated IGWs to tunnel through the subsurface convective layers. This justifies a regime of core-generated IGWs causing moderate macroturbulence for the stars with initial mass below some 12 $M_\odot$. 

Finally, the region between 12 $M_\odot$ and the dashed line allows for both regimes to co-exist. This co-aligns with the spectroscopic luminosity $\log (\mathcal{L}/\mathcal{L_{\odot}})$ around 3.5, where we can as well see a sharp turn in the distribution of the observed macroturbulent velocities towards the highest values in Fig.\ref{fig:obs_vs_predict_vmac_FeHe}.

\section{Discussion and conclusions}\label{sec:discussion_conclusions}

In this study, we investigated the physical origin of macroturbulent spectroscopic line broadening in hot massive stars by analysing a comprehensive sample of 594 stars with masses ranging from 2.5 to 80 $M_{\odot}$  across two different metallicity regimes: the Large Magellanic Cloud (LMC) and the Galaxy with $Z$=0.006 and $Z$=0.014, respectively. Our sample included 86 stars with macroturbulent broadening measured for the first time. Our study was supplied with a grid of {\sc mesa} stellar structure models to examine the properties of subsurface convective zones. These models also allowed us to explore wave propagation and the potential for IGWs to tunnel through subsurface convective layers.

To understand the underlying mechanisms behind the macroturbulent broadening, we performed several analyses. We used multivariate linear regression to identify significant predictors in both macroturbulent velocities and convective velocities predicted by 1D models. The analyses showed that in the low-mass regime 
($\log (\mathcal{L}/\mathcal{L}_{\odot})$ < 3 or $M<12M_{\odot}$), macroturbulent velocities $\Theta$ cannot be explained by the same predictors that describe the model convective velocities. At higher masses and luminosities 
($\log (\mathcal{L}/\mathcal{L}_{\odot})$ > 3 or $M>12M_{\odot}$), the observed variability in $\Theta$ may be largely explained by the same predictors and similar coefficients as the velocities occurring in the FeCZ zone.  The coefficients of all predictors in this high-mass regime show consistently similar trends with luminosity and metallicity,  which points towards a common origin of the observed macroturbulence and convective velocities in the models.  

We also conducted a Principal Component Analysis (PCA) to compare trends between the observed and model datasets. We found that the underlying structure and trends are captured similarly by the first principal component across independently analysed observed and model datasets. In the high-mass regime, the clouds of the observed and model points align in PC1-PC2 space. On the contrary, in the low-mass regime, they are misplaced, showing that model values fail to predict the observed $\Theta$. 

We interpret the consistent results of both PCA and multivariate linear regression  as strong statistical evidence that macroturbulent velocities in stars in different mass regimes are caused by different mechanisms. In stars with initial masses larger than 12 $M_{\odot}$  (or spectroscopic luminosities $\log (\mathcal{L}/\mathcal{L_{\odot}})$ > 3), macroturbulent velocities are closely related to the convective velocities in the subsurface convection zones formed due to the iron opacity bump (FeCZ), as originally proposed by 
\citet{grassitelly_2015_584}.
At the same time, our analysis revealed a clear dependence of macroturbulent velocities on metallicity. In particular, Galactic stars (higher metallicity) exhibited systematically higher macroturbulent velocities compared to stars in the LMC (lower metallicity). This result aligns well with the model predictions: sub-surface convection zones are found to develop at both Galactic and LMC metallicities, albeit the magnitude of the convective velocities is generally lower in the latter case.

Since in all our 1D models the regions of FeCZ are situated much deeper than line formation region in the photosphere, it is likely that the surface macroturbulent velocities are formed by IGWs excited at the boundary of FeCZ.  Excitation of IGWs by the FeCZ in massive stars has hardly been studied in multi-D simulations. However, Rogers et al. (in prep.; private communication) show that such a subsurface convection zone can generate low-frequency waves which reach the stellar surface, in conjunction with higher frequency IGWs generated by core convection. In 3D models computed by \citet{schultz_stochastic_2022,schultz_turbulence-supported_2023} though, the subsurface convection layers expand further than in 1D models and cause turbulent surface with velocities ranges matching observed  macroturbulent  velocities. However, such simulations are computationally demanding, hence were computed for several snapshots in time and a few values of stellar mass. More simulations would be required to draw definitive conclusions as to whether the subsurface convection could in full explain the observed macroturbulent broadening. 

For stars with initial masses below 12 $M_{\odot}$, the convective velocities in both the FeCZ and (HeCZ) are not sufficient to explain the observed macroturbulent velocities, which, although smaller than in higher mass stars, are still significant. This points to the presence of some other mechanism that is responsible for the observed line broadening in these intermediate-mass stars. This may be seen as the change in the distribution of $\Theta$ with luminosity in Fig. \ref{fig:obs_vs_predict_vmac_FeHe}.

Our analysis of wave propagation through the stellar interiors indicated that core-generated IGWs are highly unlikely to tunnel through extensive subsurface convection layers of stars with initial masses above $\sim$30 $M_{\odot}$\footnote{The exact cut-off in stellar mass is found to depend on the evolutionary stage and metallicity of the star: above 40 $M_\odot$ at ZAMS, 20 $M_\odot$ at TAMS, and 12 $M_\odot$ in the Hertzsprung gap for the Galactic metallicity, and above 60 $M_\odot$ at ZAMS, 32 $M_\odot$ at TAMS, and 12 $M_\odot$ in the Hertzsprung gap for the LMC metallicity}. The models from the same region in the HRD also have convective velocities up to 80 \kms, sufficiently large to describe the high observed macroturbulent velocities. We conclude that macroturbulent velocities are 
linked to the FeCZ properties in this high-mass regime based on statistical properties and trends in 1D models and observations, keeping in mind results from modern 2D and 3D hydrodynamical simulations \citep{schultz_synthesizing_2023,Debnath2024}. 

The region in the HRD characterised by the initial masses below some 12 $M_\odot$ generally lacks stars with high macroturbulent velocity values. At the same time, only a few stars in that region show macroturbulent velocities below 20 \kms.  In the same region, models predict (near-)zero convective velocities, as seen in Fig.\ref{fig:levels_tun_conv}, rendering the FeCZ unlikely to cause the observed macroturbulent broadening but, at the same time, allowing the core-generated IGWs to tunnel through the subsurface convective layers. This 
may allow for
a regime of core-generated IGWs causing moderate macroturbulence for the stars with initial mass below some 12 $M_\odot$. 

Lastly, the region between 12 $M_{\odot}$ and 30 $M_{\odot}$ allows IGWs to propagate to the surface and at the same time has still large FeCZ convective velocities, meaning the possible region of coexistence of the two mechanisms. More detailed 3D modelling is required to investigate the interplay of subsurface convective layers and core-generated IGWs to determine a more physical estimation of the boundary between the two discussed regimes. 

The conclusion on the existence of different regimes aligns with the recent research of the stochastic low-frequency variability from space-based photometry in SMC, LMC, and the Galaxy (Bowman et al. 2024, in prep.). The authors conclude that FeCZ plays different roles based on age, and is only crucial for evolved stars. We draw similar conclusions, but emphasise the importance of the mass range apart from the age, as for the higher masses the FeCZ is extensive from the very ZAMS. Moreover, Bowman et al. (2024, in prep.) present a convection stability window based on Rayleigh number that aligns with our region where IGWs are allowed to tunnel through FeCZ. 
Furthermore, the authors demonstrate a lack of dependence of the observed SLFV characteristics on the metallicity of the star. Assuming that the SLFV and macroturbulence phenomena originate from the same mechanism(s), the conclusion by Bowman et al. (2024, in prep.) contradicts our findings of the macroturbulent velocities being dependent on the metallicity of the star. This discrepancy leaves room for the possibility that macroturbulence and SLFV are caused (at least partially) by different physical mechanisms. 
However, we may see systematically lower macroturbulent velocities in the spectra of LMC stars due to a more evolved sample than the Galactic sample we compare it to.

Finally, we highlight the complex interplay between different physical mechanisms contributing to macroturbulence in different mass regimes, as well as limitations in the used stellar structure models. In particular, models' 1D nature and inability to account for rotating and time-dependent convection represent major uncertainties in the interpretation of the state-of-the-art (spectroscopic and photometric) observations. Furthermore, the observed sample shows a strong correlation between macroturbulence and projected rotational velocity, which we could not account for in models. Proper treatment of the interplay between these two phenomena would require development of models that incorporate both subsurface and core-generated IGWs with rotational dynamics. Such 3D rotating models could provide deeper insights into the physical processes driving macroturbulence and SLFV, and improve our understanding of the atmospheric dynamics in massive stars.

\begin{acknowledgements}
 The authors thank the anonymous referee for encouraging remarks and acknowledge fruitful discussions with Jon Sundqvist, Nico Moens, and Dwaipayan Debnath on the comparison between convective velocities from  1D MESA models and turbulent velocities in the atmospheres and outflows of their unpublished 2D simulations.
We also extend our heartfelt thanks to Prof. Dominic Bowman and Prof. Tamara Rogers for their valuable insights and stimulating discussions during our private communications that significantly enhanced quality of this work. 
The research leading to these results has received funding from the KU\,Leuven Research Council (grant C16/18/005: PARADISE), from the Research Foundation Flanders (FWO) under grant agreement G089422N, as well as from the BELgian federal Science Policy Office (BELSPO) through PRODEX grant PLATO. CA acknowledges funding from the European Research Council (ERC) under the Horizon Europe programme (Synergy Grant agreement N$^\circ$101071505: 4D-STAR). While funded by the European Union, views and opinions expressed are however those of the author(s) only and do not necessarily reflect those of the European Union or the European Research Council. Neither the European Union nor the granting authority can be held responsible for them. 

The research is based on observations collected at the European Southern Observatory under ESO programs 1104.D-0230 and 0104.A-9001. 
We would like to express our sincere gratitude to Sergio Simón-Díaz for generously providing tables with the data from \citet{simon-diaz_iacob_2017}, which were crucial for our analysis. 

\end{acknowledgements}
\bibliographystyle{aa}
\bibliography{aa}

\begin{thebibliography}{64}
\expandafter\ifx\csname natexlab\endcsname\relax\def\natexlab#1{#1}\fi

\bibitem[{{Aerts} {et~al.}(2010){Aerts}, {Christensen-Dalsgaard}, \& {Kurtz}}]{asteroseismology}
{Aerts}, C., {Christensen-Dalsgaard}, J., \& {Kurtz}, D.~W. 2010, Asteroseismology (Astronomy and Astrophysics Library. ISBN 978-1-4020-5178-4. Springer Science+Business Media)

\bibitem[{{Aerts} {et~al.}(2023){Aerts}, {Molenberghs}, \& {De Ridder}}]{Aerts2023}
{Aerts}, C., {Molenberghs}, G., \& {De Ridder}, J. 2023, \aap, 672, A183

\bibitem[{Aerts {et~al.}(2009{\natexlab{a}})Aerts, Puls, Godart, \& Dupret}]{aerts_collective_2009}
Aerts, C., Puls, J., Godart, M., \& Dupret, M.~A. 2009{\natexlab{a}}, \aap, 508, 409

\bibitem[{Aerts {et~al.}(2009{\natexlab{b}})Aerts, Puls, Godart, \& Dupret}]{aerts_origin_2009}
Aerts, C., Puls, J., Godart, M., \& Dupret, M.~A. 2009{\natexlab{b}}, Communications in Asteroseismology, 158, 66

\bibitem[{{Aerts} {et~al.}(2017){Aerts}, {S{\'\i}mon-D{\'\i}az}, {Bloemen}, {Debosscher}, {P{\'a}pics}, {Bryson}, {Still}, {Moravveji}, {Williamson}, {Grundahl}, {Fredslund Andersen}, {Antoci}, {Pall{\'e}}, {Christensen-Dalsgaard}, \& {Rogers}}]{Aerts2017a}
{Aerts}, C., {S{\'\i}mon-D{\'\i}az}, S., {Bloemen}, S., {et~al.} 2017, \aap, 602, A32

\bibitem[{Aerts {et~al.}(2014)Aerts, Simón-Díaz, Groot, \& Degroote}]{aerts_use_2014}
Aerts, C., Simón-Díaz, S., Groot, P.~J., \& Degroote, P. 2014, \aap, 569, A118

\bibitem[{Anders {et~al.}(2023)Anders, Lecoanet, Cantiello, Burns, Hyatt, Kaufman, Townsend, Brown, Vasil, Oishi, \& Jermyn}]{anders_photometric_2023}
Anders, E.~H., Lecoanet, D., Cantiello, M., {et~al.} 2023, Nature Astronomy, 7, 1228

\bibitem[{{Blomme} {et~al.}(2011){Blomme}, {Mahy}, {Catala}, {Cuypers}, {Gosset}, {Godart}, {Montalban}, {Ventura}, {Rauw}, {Morel}, {Degroote}, {Aerts}, {Noels}, {Michel}, {Baudin}, {Baglin}, {Auvergne}, \& {Samadi}}]{Blomme2011}
{Blomme}, R., {Mahy}, L., {Catala}, C., {et~al.} 2011, \aap, 533, A4

\bibitem[{Bowman {et~al.}(2019{\natexlab{a}})Bowman, Aerts, Johnston, Pedersen, Rogers, Edelmann, Simón-Díaz, Van~Reeth, Buysschaert, Tkachenko, \& Triana}]{bowman_photometric_2019}
Bowman, D.~M., Aerts, C., Johnston, C., {et~al.} 2019{\natexlab{a}}, \aap, 621, A135

\bibitem[{Bowman {et~al.}(2019{\natexlab{b}})Bowman, Burssens, Pedersen, Johnston, Aerts, Buysschaert, Michielsen, Tkachenko, Rogers, Edelmann, Ratnasingam, Simón-Díaz, Castro, Moravveji, Pope, White, \& De~Cat}]{bowman_low-frequency_2019}
Bowman, D.~M., Burssens, S., Pedersen, M.~G., {et~al.} 2019{\natexlab{b}}, Nature Astronomy, 3, 760

\bibitem[{Bowman {et~al.}(2020)Bowman, Burssens, Simón-Díaz, Edelmann, Rogers, Horst, Röpke, \& Aerts}]{bowman_photometric_2020}
Bowman, D.~M., Burssens, S., Simón-Díaz, S., {et~al.} 2020, \aap, 640, A36

\bibitem[{{Brahm} {et~al.}(2017){Brahm}, {Jord{\'a}n}, \& {Espinoza}}]{Brahm2017_ceres}
{Brahm}, R., {Jord{\'a}n}, A., \& {Espinoza}, N. 2017, \pasp, 129, 034002

\bibitem[{{Burssens} {et~al.}(2020){Burssens}, {Sim{\'o}n-D{\'\i}az}, {Bowman}, {Holgado}, {Michielsen}, {de Burgos}, {Castro}, {Barb{\'a}}, \& {Aerts}}]{Burssens2020}
{Burssens}, S., {Sim{\'o}n-D{\'\i}az}, S., {Bowman}, D.~M., {et~al.} 2020, \aap, 639, A81

\bibitem[{Cantiello {et~al.}(2009)Cantiello, Langer, Brott, de~Koter, Shore, Vink, Voegler, Lennon, \& Yoon}]{cantiello_sub-surface_2009}
Cantiello, M., Langer, N., Brott, I., {et~al.} 2009, \aap, 499, 279

\bibitem[{Cantiello {et~al.}(2021)Cantiello, Lecoanet, Jermyn, \& Grassitelli}]{cantiello_origin_2021}
Cantiello, M., Lecoanet, D., Jermyn, A.~S., \& Grassitelli, L. 2021, \apj, 915, 112

\bibitem[{{Conti} \& {Ebbets}(1977)}]{conti_ebbets_1977}
{Conti}, P.~S. \& {Ebbets}, D. 1977, \apj, 213, 438

\bibitem[{{Debnath} {et~al.}(2024){Debnath}, {Sundqvist}, {Moens}, {Van der Sijpt}, {Verhamme}, \& {Poniatowski}}]{Debnath2024}
{Debnath}, D., {Sundqvist}, J.~O., {Moens}, N., {et~al.} 2024, \aap, 684, A177

\bibitem[{{Dekker} {et~al.}(2000){Dekker}, {D'Odorico}, {Kaufer}, {Delabre}, \& {Kotzlowski}}]{uves}
{Dekker}, H., {D'Odorico}, S., {Kaufer}, A., {Delabre}, B., \& {Kotzlowski}, H. 2000, Proc. SPIE, 4008, 534

\bibitem[{Edelmann {et~al.}(2019)Edelmann, Ratnasingam, Pedersen, Bowman, Prat, \& Rogers}]{edelmann_three-dimensional_2019}
Edelmann, P. V.~F., Ratnasingam, R.~P., Pedersen, M.~G., {et~al.} 2019, \apj, 876, 4

\bibitem[{{Gebruers} {et~al.}(2022){Gebruers}, {Tkachenko}, {Bowman}, {Van Reeth}, {Burssens}, {IJspeert}, {Mahy}, {Straumit}, {Xiang}, {Rix}, \& {Aerts}}]{Gebruers2022}
{Gebruers}, S., {Tkachenko}, A., {Bowman}, D.~M., {et~al.} 2022, \aap, 665, A36

\bibitem[{{Grassitelli} {et~al.}(2015){Grassitelli}, {Fossati}, {Langer}, {Miglio}, {Istrate}, \& {Sanyal}}]{grassitelly_2015_584}
{Grassitelli}, L., {Fossati}, L., {Langer}, N., {et~al.} 2015, \aap, 584, L2

\bibitem[{Grassitelli {et~al.}(2016)Grassitelli, Fossati, Langer, Simón-Díaz, Castro, \& Sanyal}]{grassitelli_metallicity_2016}
Grassitelli, L., Fossati, L., Langer, N., {et~al.} 2016, \aap, 593, A14

\bibitem[{Grassitelli {et~al.}(2015)Grassitelli, Fossati, Simón-Diáz, Langer, Castro, \& Sanyal}]{grassitelli_observational_2015}
Grassitelli, L., Fossati, L., Simón-Diáz, S., {et~al.} 2015, \apj, 808, L31

\bibitem[{Gray(1978)}]{gray_turbulence_1978}
Gray, D.~F. 1978, \solphys, 59, 193

\bibitem[{{Gray}(2008)}]{Gray2008}
{Gray}, D.~F. 2008, {The Observation and Analysis of Stellar Photospheres}

\bibitem[{{Hillier} \& {Lanz}(2001)}]{hiller2001_cmfgen}
{Hillier}, D.~J. \& {Lanz}, T. 2001, in Astronomical Society of the Pacific Conference Series, Vol. 247, Spectroscopic Challenges of Photoionized Plasmas, ed. G.~{Ferland} \& D.~W. {Savin}, 343

\bibitem[{{Holgado} {et~al.}(2020){Holgado}, {Sim{\'o}n-D{\'\i}az}, {Haemmerl{\'e}}, {Lennon}, {Barb{\'a}}, {Cervi{\~n}o}, {Castro}, {Herrero}, {Meynet}, \& {Arias}}]{holgado2020}
{Holgado}, G., {Sim{\'o}n-D{\'\i}az}, S., {Haemmerl{\'e}}, L., {et~al.} 2020, \aap, 638, A157

\bibitem[{Howarth(2004)}]{howarth_rotation_2004}
Howarth, I.~D. 2004, 215, 33

\bibitem[{Howarth {et~al.}(1997)Howarth, Siebert, Hussain, \& Prinja}]{howarth_cross_1997}
Howarth, I.~D., Siebert, K.~W., Hussain, G. A.~J., \& Prinja, R.~K. 1997, \mnras, 284, 265

\bibitem[{{Hubeny}(1988)}]{hubeny1988_tlusty}
{Hubeny}, I. 1988, Computer Physics Communications, 52, 103

\bibitem[{{Jermyn} {et~al.}(2023){Jermyn}, {Bauer}, {Schwab}, {Farmer}, {Ball}, {Bellinger}, {Dotter}, {Joyce}, {Marchant}, {Mombarg}, {Wolf}, {Sunny Wong}, {Cinquegrana}, {Farrell}, {Smolec}, {Thoul}, {Cantiello}, {Herwig}, {Toloza}, {Bildsten}, {Townsend}, \& {Timmes}}]{Jermyn2023}
{Jermyn}, A.~S., {Bauer}, E.~B., {Schwab}, J., {et~al.} 2023, \apjs, 265, 15

\bibitem[{Kaufer {et~al.}(1999)Kaufer, Stahl, Tubbesing, Norregaard, Avila, Francois, Pasquini, \& Pizzella}]{Kaufer1999_feros}
Kaufer, A., Stahl, O., Tubbesing, S., {et~al.} 1999, The Messenger, 95, 8

\bibitem[{Kaufer {et~al.}(1997)Kaufer, Stahl, Wolf, Fullerton, Gaeng, Gummersbach, Jankovics, Kovacs, Mandel, Peitz, Rivinius, \& Szeifert}]{kaufer_long-term_1997}
Kaufer, A., Stahl, O., Wolf, B., {et~al.} 1997, \aap, 320, 273

\bibitem[{Langer \& Kudritzki(2014)}]{langer_spectroscopic_2014}
Langer, N. \& Kudritzki, R.~P. 2014, \aap, 564, A52

\bibitem[{Lecoanet {et~al.}(2021)Lecoanet, Cantiello, Anders, Quataert, Couston, Bouffard, Favier, \& Le~Bars}]{lecoanet_surface_2021}
Lecoanet, D., Cantiello, M., Anders, E.~H., {et~al.} 2021, \mnras, 508, 132

\bibitem[{Lecoanet {et~al.}(2019)Lecoanet, Cantiello, Quataert, Couston, Burns, Pope, Jermyn, Favier, \& Le~Bars}]{lecoanet_low-frequency_2019}
Lecoanet, D., Cantiello, M., Quataert, E., {et~al.} 2019, \apj, 886, L15

\bibitem[{Lucy(1976)}]{lucy_analysis_1976}
Lucy, L.~B. 1976, \apj, 206, 499

\bibitem[{Ma {et~al.}(2024)Ma, Johnston, Bellinger, \& de~Mink}]{ma_variability_2024}
Ma, L., Johnston, C., Bellinger, E.~P., \& de~Mink, S.~E. 2024, \apj, 966, 196

\bibitem[{{Markova} {et~al.}(2018){Markova}, {Puls}, \& {Langer}}]{markova2018}
{Markova}, N., {Puls}, J., \& {Langer}, N. 2018, \aap, 613, A12

\bibitem[{{Michielsen} {et~al.}(2023){Michielsen}, {Van Reeth}, {Tkachenko}, \& {Aerts}}]{Michielsen2023}
{Michielsen}, M., {Van Reeth}, T., {Tkachenko}, A., \& {Aerts}, C. 2023, \aap, 679, A6

\bibitem[{{Paxton} {et~al.}(2011){Paxton}, {Bildsten}, {Dotter}, {Herwig}, {Lesaffre}, \& {Timmes}}]{Paxton2011}
{Paxton}, B., {Bildsten}, L., {Dotter}, A., {et~al.} 2011, \apjs, 192, 3

\bibitem[{{Paxton} {et~al.}(2013){Paxton}, {Cantiello}, {Arras}, {Bildsten}, {Brown}, {Dotter}, {Mankovich}, {Montgomery}, {Stello}, {Timmes}, \& {Townsend}}]{Paxton2013}
{Paxton}, B., {Cantiello}, M., {Arras}, P., {et~al.} 2013, \apjs, 208, 4

\bibitem[{{Paxton} {et~al.}(2015){Paxton}, {Marchant}, {Schwab}, {Bauer}, {Bildsten}, {Cantiello}, {Dessart}, {Farmer}, {Hu}, {Langer}, {Townsend}, {Townsley}, \& {Timmes}}]{Paxton2015}
{Paxton}, B., {Marchant}, P., {Schwab}, J., {et~al.} 2015, \apjs, 220, 15

\bibitem[{{Paxton} {et~al.}(2018){Paxton}, {Schwab}, {Bauer}, {Bildsten}, {Blinnikov}, {Duffell}, {Farmer}, {Goldberg}, {Marchant}, {Sorokina}, {Thoul}, {Townsend}, \& {Timmes}}]{Paxton2018}
{Paxton}, B., {Schwab}, J., {Bauer}, E.~B., {et~al.} 2018, \apjs, 234, 34

\bibitem[{{Paxton} {et~al.}(2019){Paxton}, {Smolec}, {Schwab}, {Gautschy}, {Bildsten}, {Cantiello}, {Dotter}, {Farmer}, {Goldberg}, {Jermyn}, {Kanbur}, {Marchant}, {Thoul}, {Townsend}, {Wolf}, {Zhang}, \& {Timmes}}]{Paxton2019}
{Paxton}, B., {Smolec}, R., {Schwab}, J., {et~al.} 2019, \apjs, 243, 10

\bibitem[{{Puls} {et~al.}(2005){Puls}, {Urbaneja}, {Venero}, {Repolust}, {Springmann}, {Jokuthy}, \& {Mokiem}}]{puls2005_fastwind}
{Puls}, J., {Urbaneja}, M.~A., {Venero}, R., {et~al.} 2005, \aap, 435, 669

\bibitem[{{Raskin} {et~al.}(2011){Raskin}, {van Winckel}, {Hensberge}, {Jorissen}, {Lehmann}, {Waelkens}, {Avila}, {de Cuyper}, {Degroote}, {Dubosson}, {Dumortier}, {Fr{\'e}mat}, {Laux}, {Michaud}, {Morren}, {Perez Padilla}, {Pessemier}, {Prins}, {Smolders}, {van Eck}, \& {Winkler}}]{raskin2011_hermes}
{Raskin}, G., {van Winckel}, H., {Hensberge}, H., {et~al.} 2011, \aap, 526, A69

\bibitem[{Ratnasingam {et~al.}(2020)Ratnasingam, Edelmann, \& Rogers}]{ratnasingam_two-dimensional_2020}
Ratnasingam, R.~P., Edelmann, P. V.~F., \& Rogers, T.~M. 2020, \mnras, 497, 4231

\bibitem[{Ratnasingam {et~al.}(2023)Ratnasingam, Rogers, Chowdhury, Handler, Vanon, Varghese, \& Edelmann}]{ratnasingam_internal_2023}
Ratnasingam, R.~P., Rogers, T.~M., Chowdhury, S., {et~al.} 2023, \aap, 674, A134

\bibitem[{Rogers {et~al.}(2013)Rogers, Lin, McElwaine, \& Lau}]{rogers_internal_2013}
Rogers, T.~M., Lin, D. N.~C., McElwaine, J.~N., \& Lau, H. H.~B. 2013, \apj, 772, 21

\bibitem[{Ryans {et~al.}(2002)Ryans, Dufton, Rolleston, Lennon, Keenan, Smoker, \& Lambert}]{ryans_macroturbulent_2002}
Ryans, R. S.~I., Dufton, P.~L., Rolleston, W. R.~J., {et~al.} 2002, \mnras, 336, 577

\bibitem[{Schultz {et~al.}(2022)Schultz, Bildsten, \& Jiang}]{schultz_stochastic_2022}
Schultz, W.~C., Bildsten, L., \& Jiang, Y.-F. 2022, \apj, 924, L11

\bibitem[{Schultz {et~al.}(2023{\natexlab{a}})Schultz, Bildsten, \& Jiang}]{schultz_turbulence-supported_2023}
Schultz, W.~C., Bildsten, L., \& Jiang, Y.-F. 2023{\natexlab{a}}, \apj, 951, L42

\bibitem[{Schultz {et~al.}(2023{\natexlab{b}})Schultz, Tsang, Bildsten, \& Jiang}]{schultz_synthesizing_2023}
Schultz, W.~C., Tsang, B. T.~H., Bildsten, L., \& Jiang, Y.-F. 2023{\natexlab{b}}, \apj, 945, 58

\bibitem[{{Serebriakova} {et~al.}(2023){Serebriakova}, {Tkachenko}, {Gebruers}, {Bowman}, {Van Reeth}, {Mahy}, {Burssens}, {IJspeert}, {Sana}, \& {Aerts}}]{Serebriakova2023}
{Serebriakova}, N., {Tkachenko}, A., {Gebruers}, S., {et~al.} 2023, \aap, 676, A85

\bibitem[{{Sim{\'o}n-D{\'\i}az} {et~al.}(2011){Sim{\'o}n-D{\'\i}az}, {Castro}, {Garcia}, {Herrero}, \& {Markova}}]{diaz2011_iacob}
{Sim{\'o}n-D{\'\i}az}, S., {Castro}, N., {Garcia}, M., {Herrero}, A., \& {Markova}, N. 2011, Bulletin de la Societe Royale des Sciences de Liege, 80, 514

\bibitem[{{Sim{\'o}n-D{\'\i}az} {et~al.}(2017){Sim{\'o}n-D{\'\i}az}, {Godart}, {Castro}, {Herrero}, {Aerts}, {Puls}, {Telting}, \& {Grassitelli}}]{Simon-Diaz2017}
{Sim{\'o}n-D{\'\i}az}, S., {Godart}, M., {Castro}, N., {et~al.} 2017, \aap, 597, A22

\bibitem[{Simón-Díaz {et~al.}(2017)Simón-Díaz, Godart, Castro, Herrero, Aerts, Puls, Telting, \& Grassitelli}]{simon-diaz_iacob_2017}
Simón-Díaz, S., Godart, M., Castro, N., {et~al.} 2017, \aap, 597, A22

\bibitem[{Simón-Díaz {et~al.}(2010)Simón-Díaz, Herrero, Uytterhoeven, Castro, Aerts, \& Puls}]{simon-diaz_observational_2010}
Simón-Díaz, S., Herrero, A., Uytterhoeven, K., {et~al.} 2010, \apj, 720, L174

\bibitem[{{Straumit} {et~al.}(2022){Straumit}, {Tkachenko}, {Gebruers}, {Audenaert}, {Xiang}, {Zari}, {Aerts}, {Johnson}, {Kollmeier}, {Rix}, {Beaton}, {Van Saders}, {Teske}, {Roman-Lopes}, {Ting}, \& {Rom{\'a}n-Z{\'u}{\~n}iga}}]{straumit2022_zetapayne}
{Straumit}, I., {Tkachenko}, A., {Gebruers}, S., {et~al.} 2022, \aj, 163, 236

\bibitem[{{Telting} {et~al.}(2014){Telting}, {Avila}, {Buchhave}, {Frandsen}, {Gandolfi}, {Lindberg}, {Stempels}, {Prins}, \& {NOT staff}}]{telting2014_fies}
{Telting}, J.~H., {Avila}, G., {Buchhave}, L., {et~al.} 2014, Astronomische Nachrichten, 335, 41

\bibitem[{{Tkachenko}(2015)}]{tkachenko2015_gssp}
{Tkachenko}, A. 2015, \aap, 581, A129

\bibitem[{{Tkachenko} {et~al.}(2014){Tkachenko}, {Degroote}, {Aerts}, {Pavlovski}, {Southworth}, {P{\'a}pics}, {Moravveji}, {Kolbas}, {Tsymbal}, {Debosscher}, \& {Cl{\'e}mer}}]{Tkachenko2014}
{Tkachenko}, A., {Degroote}, P., {Aerts}, C., {et~al.} 2014, \mnras, 438, 3093

\bibitem[{Vanon {et~al.}(2023)Vanon, Edelmann, Ratnasingam, Varghese, \& Rogers}]{vanon_three-dimensional_2023}
Vanon, R., Edelmann, P. V.~F., Ratnasingam, R.~P., Varghese, A., \& Rogers, T.~M. 2023, \apj, 954, 171

\end{thebibliography}

\appendix

\onecolumn

\section{Kippenhahn diagrams for LMC}
\label{app:kippenhahn}

\begin{figure*}[!htbp]
   \begin{centering}
            {\includegraphics[clip,width=540pt,trim={0.0cm 0.0cm 0.0cm 0.0cm}]{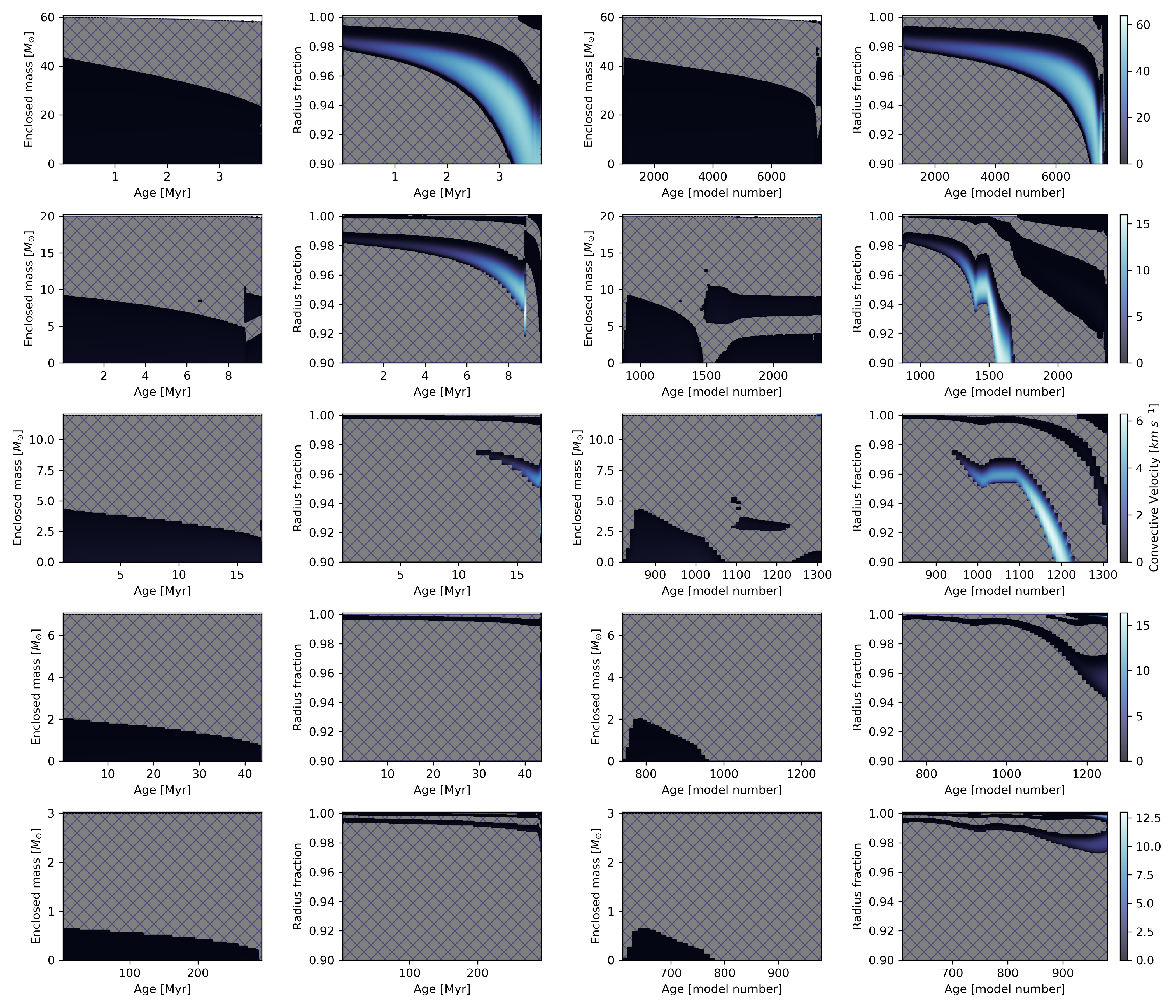}}
            
      \caption{ Kippenhahn diagrams for 60, 20, 12, 7, and 3 $M_{\odot}$ (rows) for metallicity $Z$=0.006. Odd columns -- full radius (in mass units), even columns -- zoom into outer 10\% of radius (in radius units). Hatched areas denote radiative zones. The Ledoux-unstable convective zones are colour-mapped based on convective velocity. Two left columns are plotted with age in Myr, while two right columns are plotted with age in model number, which allows a better view of the short-lived post-MS stages.}
         \label{fig:kippenhahn006}
   \end{centering}
\end{figure*}

\clearpage

\section{Full tables with results of multivariate linear regression}
\label{app:tables}

\begin{table*} [!htbp]
\centering
\caption{Multivariate linear regression: maximum model convective velocities}

\begin{tabular}{lcccccc}
\toprule
Dataset & $\log (\mathcal{L}/\mathcal{L}_{\odot})$ & log\teff & \logg & \text{Const} & \textbf{Adj. $R^2$} \\
\midrule 

\multirow{2}{*}{$Z=0.014,  v_{\text{conv, Fe}}^{\text{max}}, \text{all} \log (\mathcal{L}/\mathcal{L}_{\odot})$ }  & \multirow{2}{*}{0.000} & \multirow{2}{*}{0.000} & \multirow{2}{*}{0.000} & \multirow{2}{*}{0.000} & \multirow{2}{*}{\textbf{0.681}} \\ & 109.69 & -99.78 & 37.62 & -51.47 & \\

\multirow{2}{*}{$Z=0.014,  v_{\text{conv, Fe}}^{\text{max}}, \log (\mathcal{L}/\mathcal{L}_{\odot}) > 2.5$} & \multirow{2}{*}{0.000} & \multirow{2}{*}{0.000} & \multirow{2}{*}{0.000} & \multirow{2}{*}{0.000} & \multirow{2}{*}{\textbf{0.718}} \\ & 118.62 & -109.03 & 40.73 & -56.09 & \\
 
\multirow{2}{*}{$Z=0.014,  v_{\text{conv, Fe}}^{\text{max}}, \log (\mathcal{L}/\mathcal{L}_{\odot}) > 3.0$} & \multirow{2}{*}{0.000} & \multirow{2}{*}{0.000} & \multirow{2}{*}{0.000} & \multirow{2}{*}{0.000} & \multirow{2}{*}{\textbf{0.794}} \\ & 155.85 & -148.63 & 54.81 & -75.84 & \\
 
\multirow{2}{*}{$Z=0.014,  v_{\text{conv, Fe}}^{\text{max}}, \log (\mathcal{L}/\mathcal{L}_{\odot}) <2.5$} & \multirow{2}{*}{0.000} & \multirow{2}{*}{0.017} & \multirow{2}{*}{0.009} & \multirow{2}{*}{0.005} & \multirow{2}{*}{\textbf{0.190}} \\ & 2.14 & -1.08 & 0.75 & -0.68  \\ \hline

\multirow{2}{*}{$Z=0.006,  v_{\text{conv, Fe}}^{\text{max}}, \text{all} \log (\mathcal{L}/\mathcal{L}_{\odot})$ }   & \multirow{2}{*}{0.000} & \multirow{2}{*}{0.000} & \multirow{2}{*}{0.000} & \multirow{2}{*}{0.000} & \multirow{2}{*}{\textbf{0.460}} \\ & 58.79 & -51.03 & 16.16 & -26.28 & \\
 
\multirow{2}{*}{$Z=0.006,  v_{\text{conv, Fe}}^{\text{max}}, \log (\mathcal{L}/\mathcal{L}_{\odot}) > 3.0$} & \multirow{2}{*}{0.000} & \multirow{2}{*}{0.000} & \multirow{2}{*}{0.000} & \multirow{2}{*}{0.000} & \multirow{2}{*}{\textbf{0.722}} \\ & 106.82 & -97.93 & 29.01 & -49.69 & \\

\multirow{2}{*}{$Z=0.006,  v_{\text{conv, Fe}}^{\text{max}}, \log (\mathcal{L}/\mathcal{L}_{\odot}) < 3.0$} & \multirow{2}{*}{0.797} & \multirow{2}{*}{0.894} & \multirow{2}{*}{0.863} & \multirow{2}{*}{0.942} & \multirow{2}{*}{\textbf{0.090}} \\ & 0.93 & 0.48 & -0.55 & 0.14 & \\ \hline


\multirow{2}{*}{$Z=0.014,  v_{\text{conv, He}}^{\text{max}}, \text{all} \log (\mathcal{L}/\mathcal{L}_{\odot})$ }  & \multirow{2}{*}{0.000} & \multirow{2}{*}{0.000} & \multirow{2}{*}{0.000} & \multirow{2}{*}{0.000} & \multirow{2}{*}{\textbf{0.422}} \\ & -3.31 & 4.72 & -3.21 & 2.39 & \\

\multirow{2}{*}{$Z=0.014,  v_{\text{conv, He}}^{\text{max}}, \log (\mathcal{L}/\mathcal{L}_{\odot}) > 2.5$} & \multirow{2}{*}{0.000} & \multirow{2}{*}{0.000} & \multirow{2}{*}{0.000} & \multirow{2}{*}{0.000} & \multirow{2}{*}{\textbf{0.230}} \\ & -1.41 & 2.32 & -1.79 & 1.17 & \\
 
\multirow{2}{*}{$Z=0.014,  v_{\text{conv, He}}^{\text{max}}, \log (\mathcal{L}/\mathcal{L}_{\odot}) > 3.0$} & \multirow{2}{*}{0.009} & \multirow{2}{*}{0.000} & \multirow{2}{*}{0.000} & \multirow{2}{*}{0.000} & \multirow{2}{*}{\textbf{0.250}} \\ & -0.36 & 1.03 & -1.07 & 0.52 & \\ 
 
\multirow{2}{*}{$Z=0.014,  v_{\text{conv, He}}^{\text{max}}, \log (\mathcal{L}/\mathcal{L}_{\odot}) < 2.5$} & \multirow{2}{*}{0.000} & \multirow{2}{*}{0.000} & \multirow{2}{*}{0.000} & \multirow{2}{*}{0.000} & \multirow{2}{*}{\textbf{0.645}} \\ & -9.23 & 12.46 & -9.83 & 6.49 & \\ \hline

\multirow{2}{*}{$Z=0.006,  v_{\text{conv, He}}^{\text{max}}, \text{all} \log (\mathcal{L}/\mathcal{L}_{\odot})$ }   & \multirow{2}{*}{0.000} & \multirow{2}{*}{0.000} & \multirow{2}{*}{0.000} & \multirow{2}{*}{0.000} & \multirow{2}{*}{\textbf{0.365}} \\ & -3.26 & 4.23 & -2.71 & 2.16 & \\
 
\multirow{2}{*}{$Z=0.006,  v_{\text{conv, He}}^{\text{max}}, \log (\mathcal{L}/\mathcal{L}_{\odot}) > 3.0$}  & \multirow{2}{*}{0.714} & \multirow{2}{*}{0.005} & \multirow{2}{*}{0.000} & \multirow{2}{*}{0.005} & \multirow{2}{*}{\textbf{0.263}} \\ & 0.04 & 0.33 & -0.52 & 0.17 & \\
 
\multirow{2}{*}{$Z=0.006,  v_{\text{conv, He}}^{\text{max}}, \log (\mathcal{L}/\mathcal{L}_{\odot}) < 3.0$}  & \multirow{2}{*}{0.000} & \multirow{2}{*}{0.000} & \multirow{2}{*}{0.000} & \multirow{2}{*}{0.000} & \multirow{2}{*}{\textbf{0.596}} \\ & -5.70 & 9.46 & -7.87 & 4.84 & \\

\bottomrule
\end{tabular}
\tablefoot{
$P>|t|$ and $c_i$ with $\log (\mathcal{L}/\mathcal{L_{\odot}})$, log\teff, and \logg~ used as predictors, and various subsets of maximum model convective velocities as dependent variables (metallicities $Z$=0.014 and $Z$=0.006, iron-bump and helium-bump convection layers $v_{\text{conv, Fe}}$ and $v_{\text{conv, He}}$, and whether cutoff in $\mathcal{L}$ to follow observations was introduced or not).
}
\label{tab:multivar_max_all}
\end{table*}

\begin{table*}[!htbp]
\centering
\caption{Multivariate linear regression: integrated model convective velocities. }
\begin{tabular}{lcccccc}
\toprule
Dataset & $\log (\mathcal{L}/\mathcal{L}_{\odot})$ & log\teff & \logg & \text{Const} & \textbf{Adj. $R^2$} \\
\midrule

\multirow{2}{*}{$Z=0.014,  v_{\text{conv, Fe}}^{\text{integr}}, \text{all} \log (\mathcal{L}/\mathcal{L}_{\odot})$ }  & \multirow{2}{*}{0.000} & \multirow{2}{*}{0.000} & \multirow{2}{*}{0.000} & \multirow{2}{*}{0.000} & \multirow{2}{*}{\textbf{0.636}} \\ & 77.04 & -66.76 & 21.29 & -34.41 & \\

\multirow{2}{*}{$Z=0.014,  v_{\text{conv, Fe}}^{\text{integr}}, \log (\mathcal{L}/\mathcal{L}_{\odot}) > 2.5$} & \multirow{2}{*}{0.000} & \multirow{2}{*}{0.000} & \multirow{2}{*}{0.000} & \multirow{2}{*}{0.000} & \multirow{2}{*}{\textbf{0.670}} \\ & 83.36 & -73.30 & 23.49 & -37.68 & \\

\multirow{2}{*}{$Z=0.014,  v_{\text{conv, Fe}}^{\text{integr}}, \log (\mathcal{L}/\mathcal{L}_{\odot}) > 3.0$}  & \multirow{2}{*}{0.000} & \multirow{2}{*}{0.000} & \multirow{2}{*}{0.000} & \multirow{2}{*}{0.000} & \multirow{2}{*}{\textbf{0.751}} \\ & 111.24 & -102.73 & 33.74 & -52.36 & \\
 
\multirow{2}{*}{$Z=0.014,  v_{\text{conv, Fe}}^{\text{integr}}, \log (\mathcal{L}/\mathcal{L}_{\odot}) <2.5$} & \multirow{2}{*}{--} & \multirow{2}{*}{--} & \multirow{2}{*}{--} & \multirow{2}{*}{--} & \multirow{2}{*}{\textbf{0}} \\ & -- & -- & -- & --  \\ \hline

\multirow{2}{*}{$Z=0.006,  v_{\text{conv, Fe}}^{\text{integr}}, \text{all} \log (\mathcal{L}/\mathcal{L}_{\odot})$ }    & \multirow{2}{*}{0.000} & \multirow{2}{*}{0.000} & \multirow{2}{*}{0.000} & \multirow{2}{*}{0.000} & \multirow{2}{*}{\textbf{0.426}} \\ & 35.16 & -28.55 & 6.81 & -14.71 & \\
 
\multirow{2}{*}{$Z=0.006,  v_{\text{conv, Fe}}^{\text{integr}}, \log (\mathcal{L}/\mathcal{L}_{\odot}) > 3.0$}  & \multirow{2}{*}{0.000} & \multirow{2}{*}{0.000} & \multirow{2}{*}{0.000} & \multirow{2}{*}{0.000} & \multirow{2}{*}{\textbf{0.667}} \\ & 63.98 & -56.70 & 14.52 & -28.76 & \\

\multirow{2}{*}{$Z=0.006,  v_{\text{conv, Fe}}^{\text{integr}}, \log (\mathcal{L}/\mathcal{L}_{\odot}) < 3.0$} & \multirow{2}{*}{--} & \multirow{2}{*}{--} & \multirow{2}{*}{--} & \multirow{2}{*}{--} & \multirow{2}{*}{\textbf{0}} \\ & -- & -- & -- & --  \\ \hline


\multirow{2}{*}{$Z=0.014,  v_{\text{conv, He}}^{\text{integr}}, \text{all} \log (\mathcal{L}/\mathcal{L}_{\odot})$ }   & \multirow{2}{*}{0.000} & \multirow{2}{*}{0.000} & \multirow{2}{*}{0.000} & \multirow{2}{*}{0.000} & \multirow{2}{*}{\textbf{0.369}} \\ & -0.84 & 1.20 & -0.81 & 0.61 & \\

\multirow{2}{*}{$Z=0.014,  v_{\text{conv, He}}^{\text{integr}}, \log (\mathcal{L}/\mathcal{L}_{\odot}) > 2.5$}  & \multirow{2}{*}{0.000} & \multirow{2}{*}{0.000} & \multirow{2}{*}{0.000} & \multirow{2}{*}{0.000} & \multirow{2}{*}{\textbf{0.241}} \\ & -0.18 & 0.44 & -0.43 & 0.22 & \\
 
\multirow{2}{*}{$Z=0.014,  v_{\text{conv, He}}^{\text{integr}}, \log (\mathcal{L}/\mathcal{L}_{\odot}) > 3.0$}  & \multirow{2}{*}{0.016} & \multirow{2}{*}{0.000} & \multirow{2}{*}{0.000} & \multirow{2}{*}{0.000} & \multirow{2}{*}{\textbf{0.234}} \\ & -0.14 & 0.41 & -0.43 & 0.21 & \\
 
\multirow{2}{*}{$Z=0.014,  v_{\text{conv, He}}^{\text{integr}}, \log (\mathcal{L}/\mathcal{L}_{\odot}) < 2.5$}  & \multirow{2}{*}{0.000} & \multirow{2}{*}{0.000} & \multirow{2}{*}{0.000} & \multirow{2}{*}{0.000} & \multirow{2}{*}{\textbf{0.502}} \\ & -2.50 & 2.86 & -2.04 & 1.50 & \\ \hline

\multirow{2}{*}{$Z=0.006,  v_{\text{conv, He}}^{\text{integr}}, \text{all} \log (\mathcal{L}/\mathcal{L}_{\odot})$ }   & \multirow{2}{*}{0.000} & \multirow{2}{*}{0.000} & \multirow{2}{*}{0.000} & \multirow{2}{*}{0.000} & \multirow{2}{*}{\textbf{0.374}} \\ & -0.82 & 1.05 & -0.66 & 0.54 & \\
 
\multirow{2}{*}{$Z=0.006,  v_{\text{conv, He}}^{\text{integr}}, \log (\mathcal{L}/\mathcal{L}_{\odot}) > 3.0$}   & \multirow{2}{*}{0.962} & \multirow{2}{*}{0.002} & \multirow{2}{*}{0.000} & \multirow{2}{*}{0.002} & \multirow{2}{*}{\textbf{0.255}} \\ & 0.00 & 0.17 & -0.24 & 0.09 & \\
 
\multirow{2}{*}{$Z=0.006,  v_{\text{conv, He}}^{\text{integr}}, \log (\mathcal{L}/\mathcal{L}_{\odot}) < 3.0$}  
 & \multirow{2}{*}{0.000} & \multirow{2}{*}{0.000} & \multirow{2}{*}{0.000} & \multirow{2}{*}{0.000} & \multirow{2}{*}{\textbf{0.489}} \\ & -1.65 & 1.93 & -1.30 & 1.01 & \\

\bottomrule
\end{tabular}
\tablefoot{$P>|t|$ and $c_i$ with $\log (\mathcal{L}/\mathcal{L_{\odot}})$, log\teff, and \logg~ as predictors, and various subsets of integrated model convective velocities as dependent variables (metallicities $Z$=0.014 and $Z$=0.006, iron-bump and helium-bump convection layers $v_{\text{conv, Fe}}$ and $v_{\text{conv, He}}$, and whether a cutoff in $\mathcal{L}$ for the observations was introduced or not).}
\label{tab:multivar_integr_all}
\end{table*}

\clearpage

\section{MESA internal profiles}
\label{app:mesaprofiles}

\begin{figure*}[!htbp]
   \begin{centering}
            {\includegraphics[clip,width=480pt,trim={0.5cm 1.0cm 0.0cm 1.5cm}]{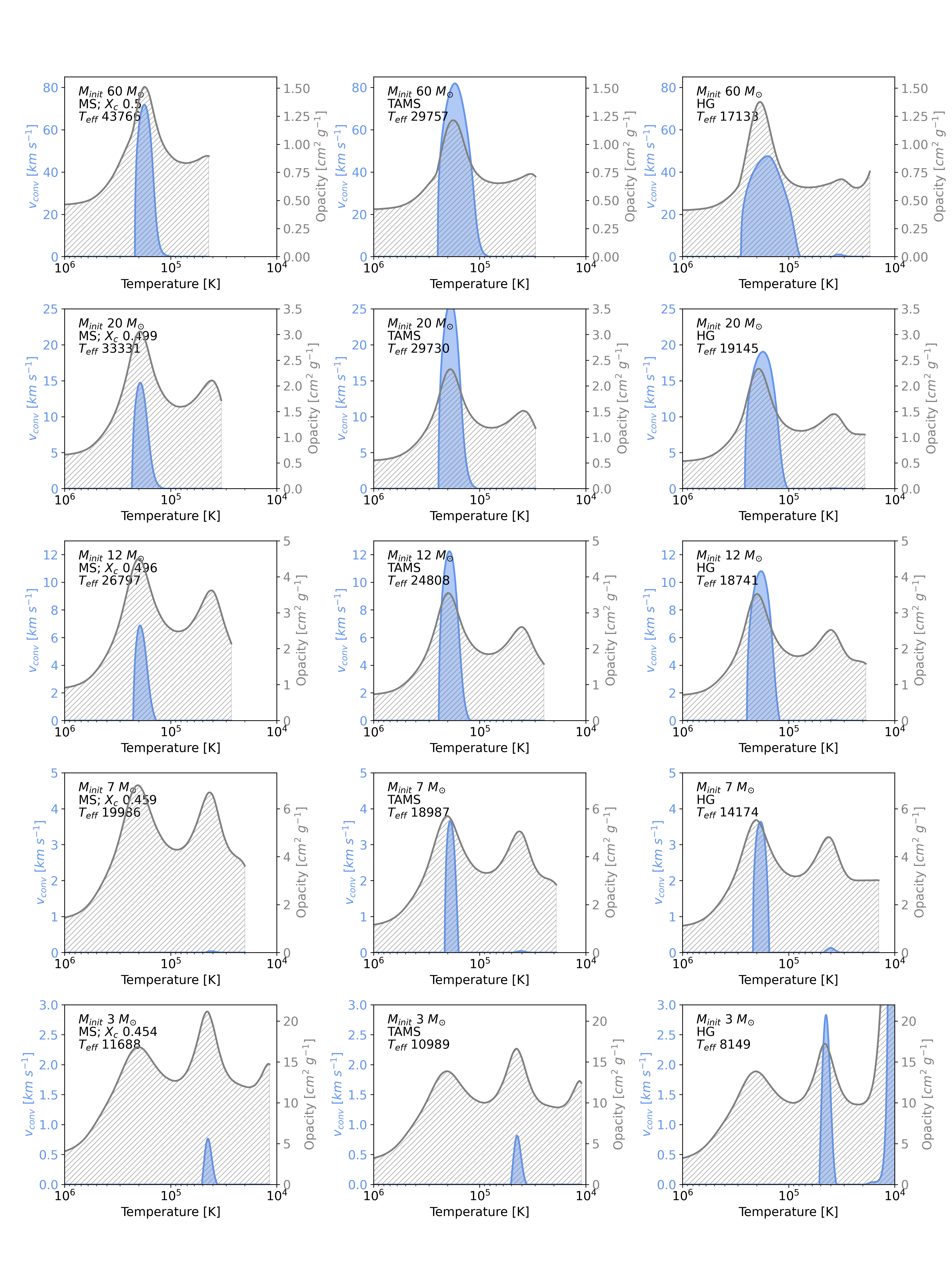}}
            
      \caption{ Profiles of opacity (grey) and convective velocity (blue) as a function of temperature. Initial mass varies in rows: 60, 20, 12, 7, and 3 $M_{\odot}$. In columns, three evolutionary stages are presented: MS, TAMS, and Hertzsprung gap. All models are computed for the Solar metallicity $Z$=0.014. }
         \label{fig:z014_opacity_convvel_vs_temperature}
   \end{centering}
\end{figure*}

\begin{figure*}
   \begin{centering}
            {\includegraphics[clip,width=480pt,trim={0.5cm 1.0cm 0.0cm 1.5cm}]{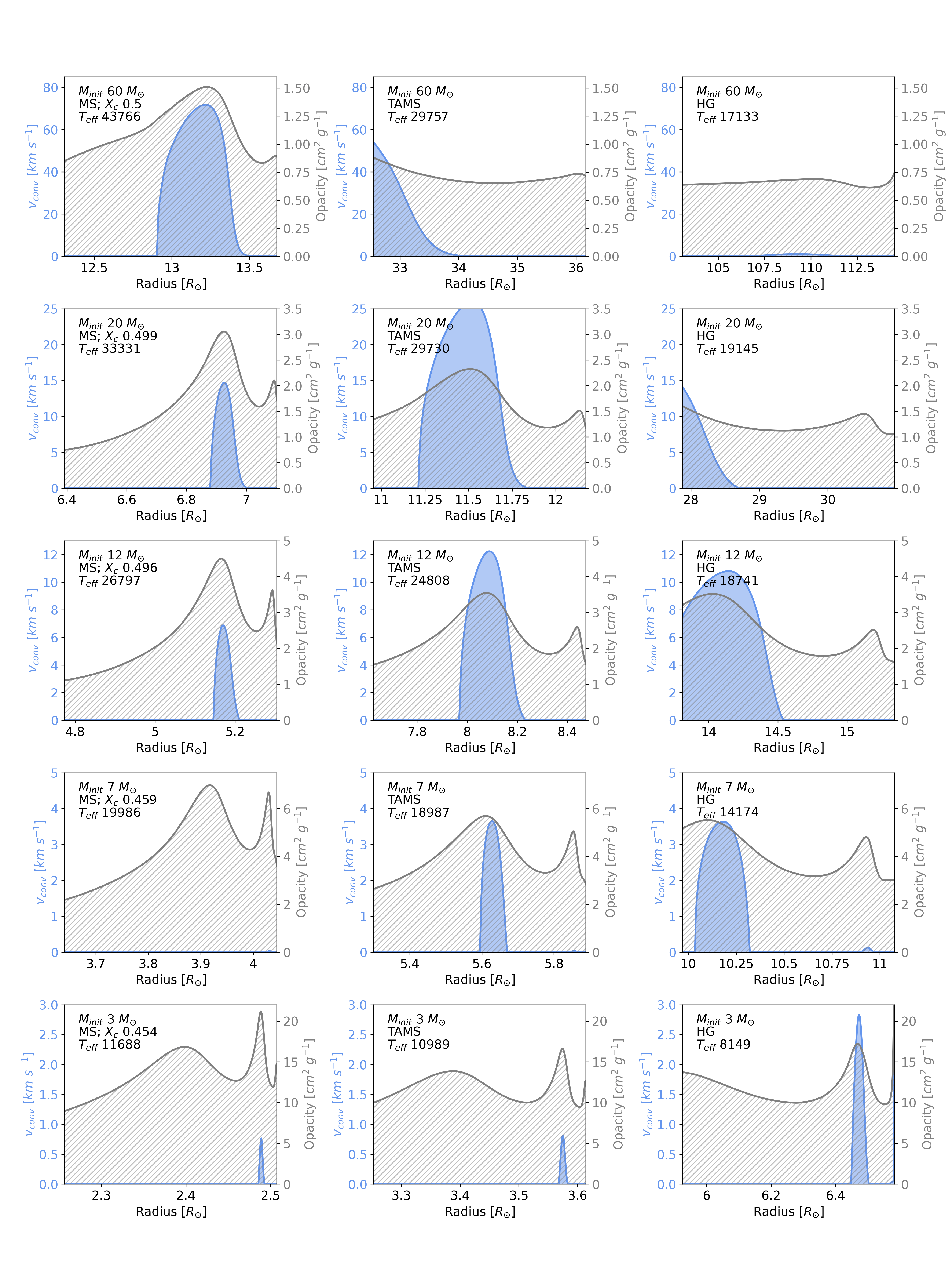}}
            
      \caption{ Profiles of opacity (grey) and convective velocity (blue) as a function of radius (zoomed into outer 10\% of the radius). The panels are the same as in Fig.\ref{fig:z014_opacity_convvel_vs_temperature}.  }
         \label{fig:z014_opacity_convvel_vs_radius_zoom}
   \end{centering}
\end{figure*}

\begin{figure*}
   \begin{centering}
            {\includegraphics[clip,width=480pt,trim={0.5cm 1.0cm 0.0cm 1.5cm}]{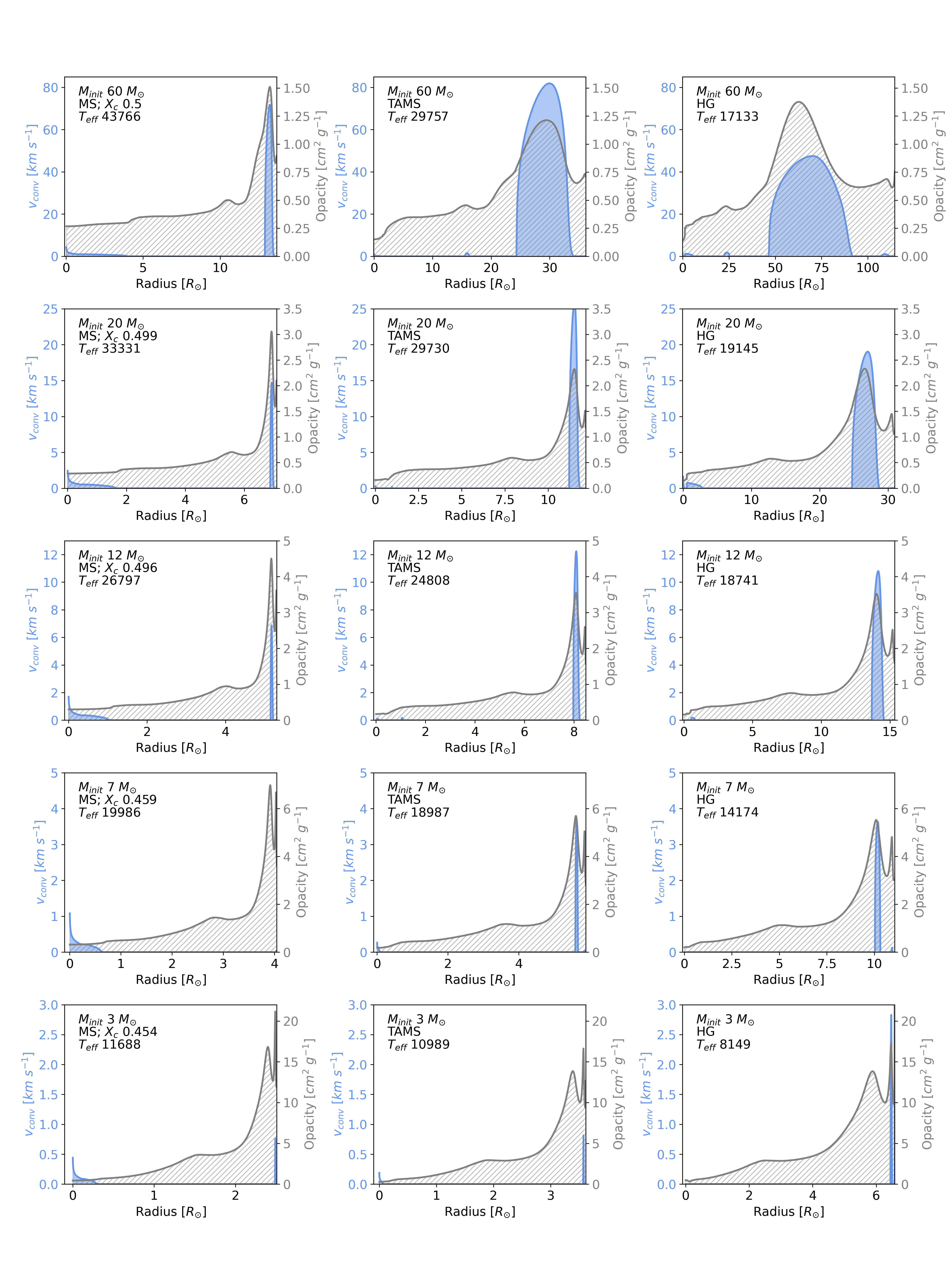}}
            
      \caption{ Profiles of opacity (grey) and convective velocity (blue) as a function of radius (full radius). The panels are the same as in Fig.\ref{fig:z014_opacity_convvel_vs_temperature}. }
         \label{fig:prof_z014_opacity_convvel_vs_radius}
   \end{centering}
\end{figure*}

\begin{figure*}
   \begin{centering}
            {\includegraphics[clip,width=480pt,trim={0.5cm 1.0cm 0.0cm 1.5cm}]{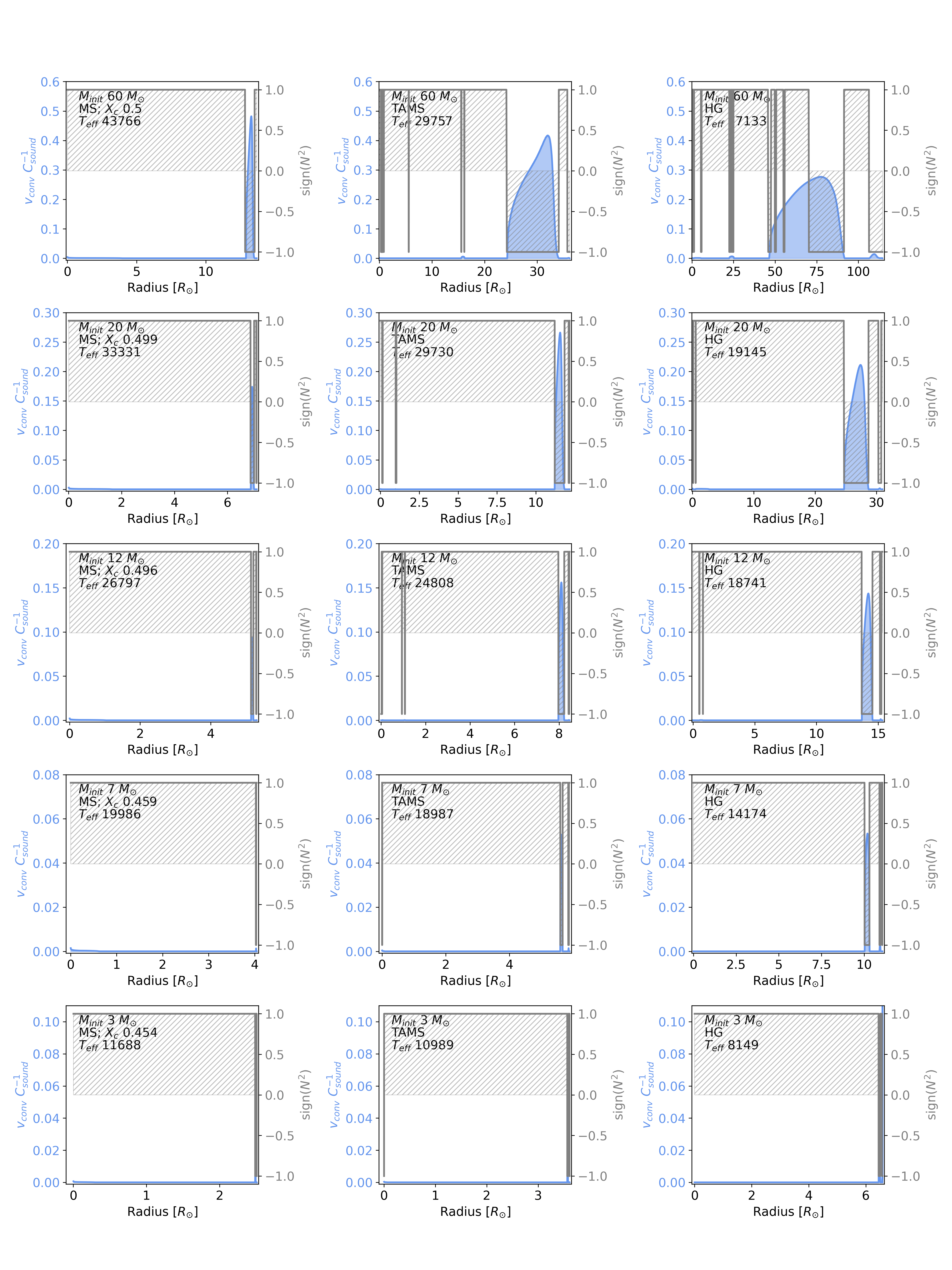}}
            
      \caption{ Profiles of the sign of the squared Brunt-Väisälä frequency (grey) and convective velocity as a fraction of sound speed (blue) as a function of radius (full radius). The panels are the same as in Fig.\ref{fig:z014_opacity_convvel_vs_temperature}. }
         \label{fig:prof_z014_csound_convvel_vs_radius}
   \end{centering}
\end{figure*}


\begin{figure*}
   \begin{centering}
            {\includegraphics[clip,width=480pt,trim={0.5cm 1.0cm 0.0cm 1.5cm}]{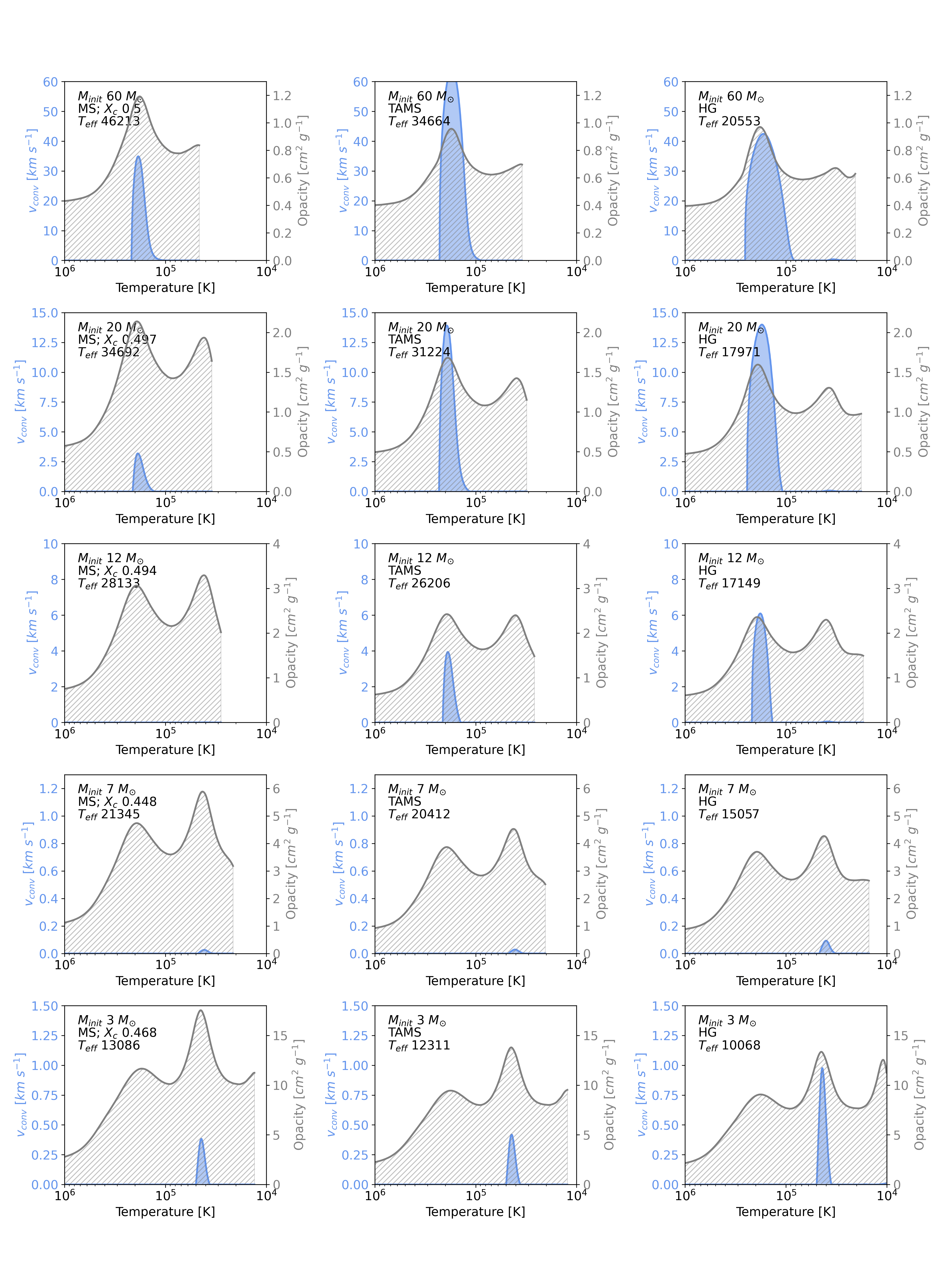}}
            
      \caption{ Profiles of opacity (grey) and convective velocity (blue) as a function of temperature. The panels are the same as in Fig.\ref{fig:z014_opacity_convvel_vs_temperature} but for $Z$=0.006. }
         \label{fig:prof_z006_opacity_convvel_vs_temperature}
   \end{centering}
\end{figure*}

\begin{figure*}
   \begin{centering}
            {\includegraphics[clip,width=480pt,trim={0.5cm 1.0cm 0.0cm 1.5cm}]{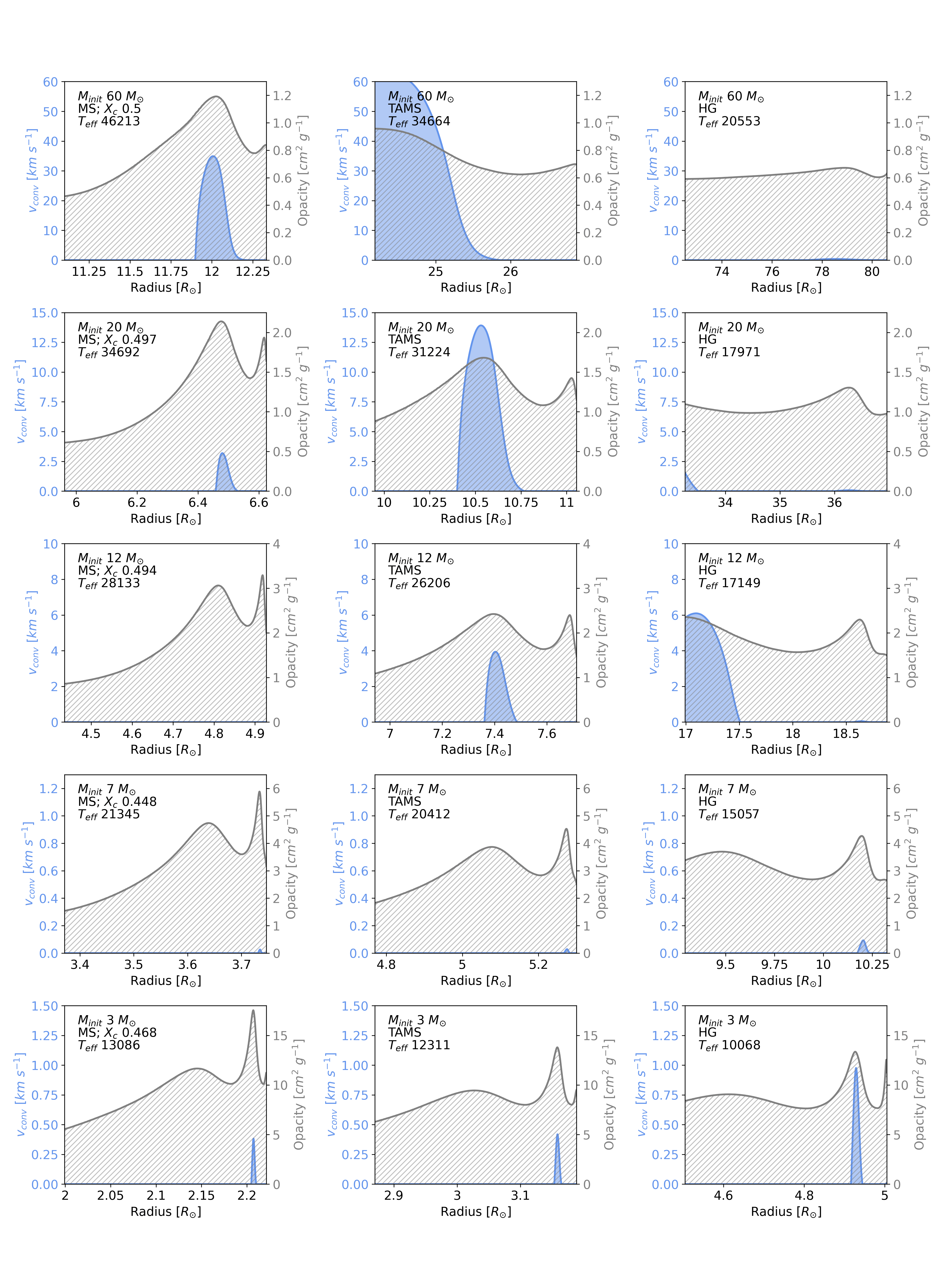}}
            
      \caption{ Profiles of opacity (grey) and convective velocity (blue) as a function of radius (zoomed into outer 10\% of radius). The panels are the same as in Fig.\ref{fig:z014_opacity_convvel_vs_temperature} but for $Z$=0.006. }
         \label{fig:prof_z006_opacity_convvel_vs_radius_zoom}
   \end{centering}
\end{figure*}

\begin{figure*}
   \begin{centering}
            {\includegraphics[clip,width=480pt,trim={0.5cm 1.0cm 0.0cm 1.5cm}]{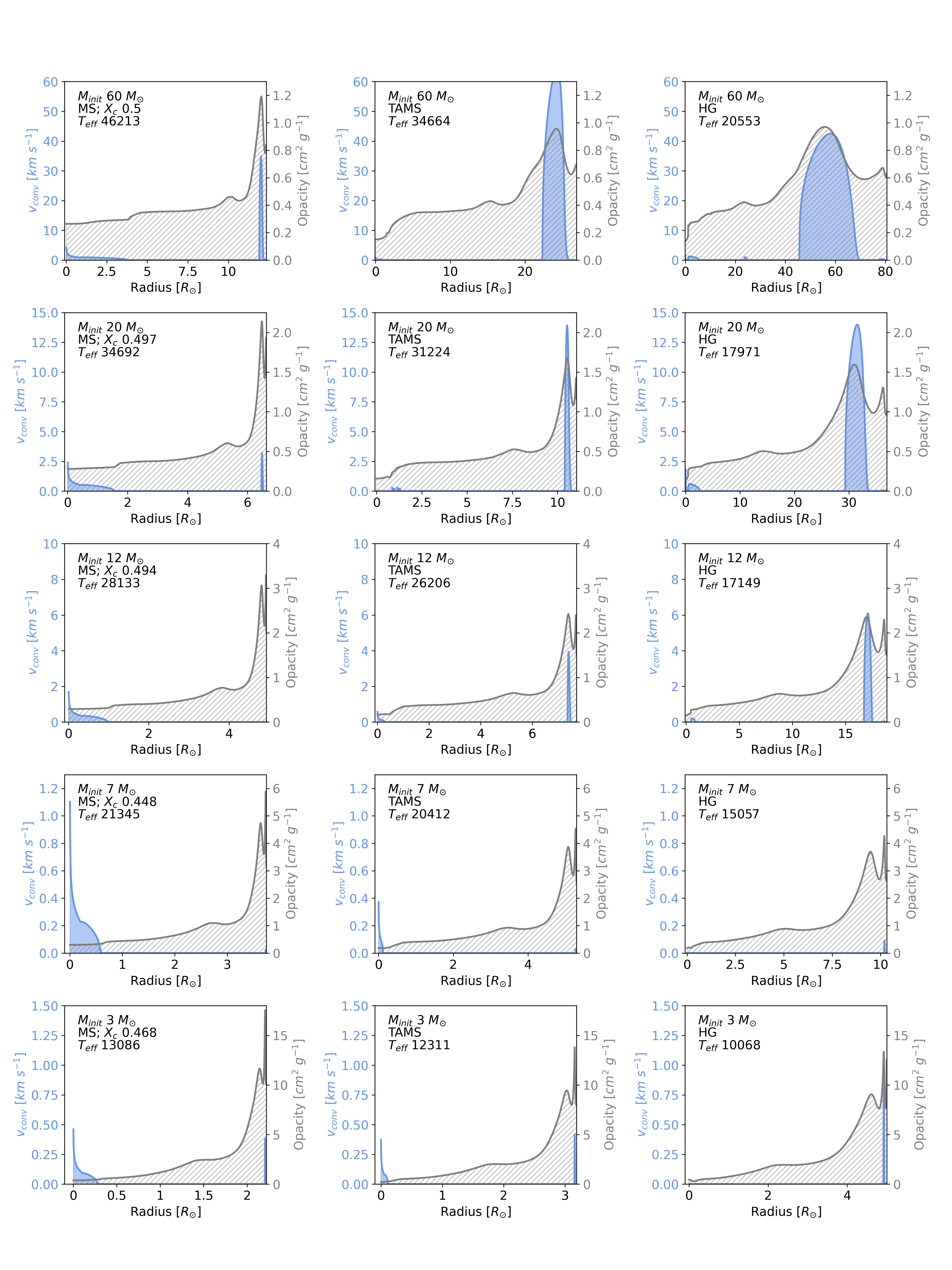}}
            
      \caption{ Profiles of opacity (grey) and convective velocity (blue) as a function of radius (full radius). The panels are the same as in Fig.\ref{fig:z014_opacity_convvel_vs_temperature} but for $Z$=0.006. }
         \label{fig:prof_z006_opacity_convvel_vs_radius}
   \end{centering}
\end{figure*}

\begin{figure*}
   \begin{centering}
            {\includegraphics[clip,width=480pt,trim={0.5cm 1.0cm 0.0cm 1.5cm}]{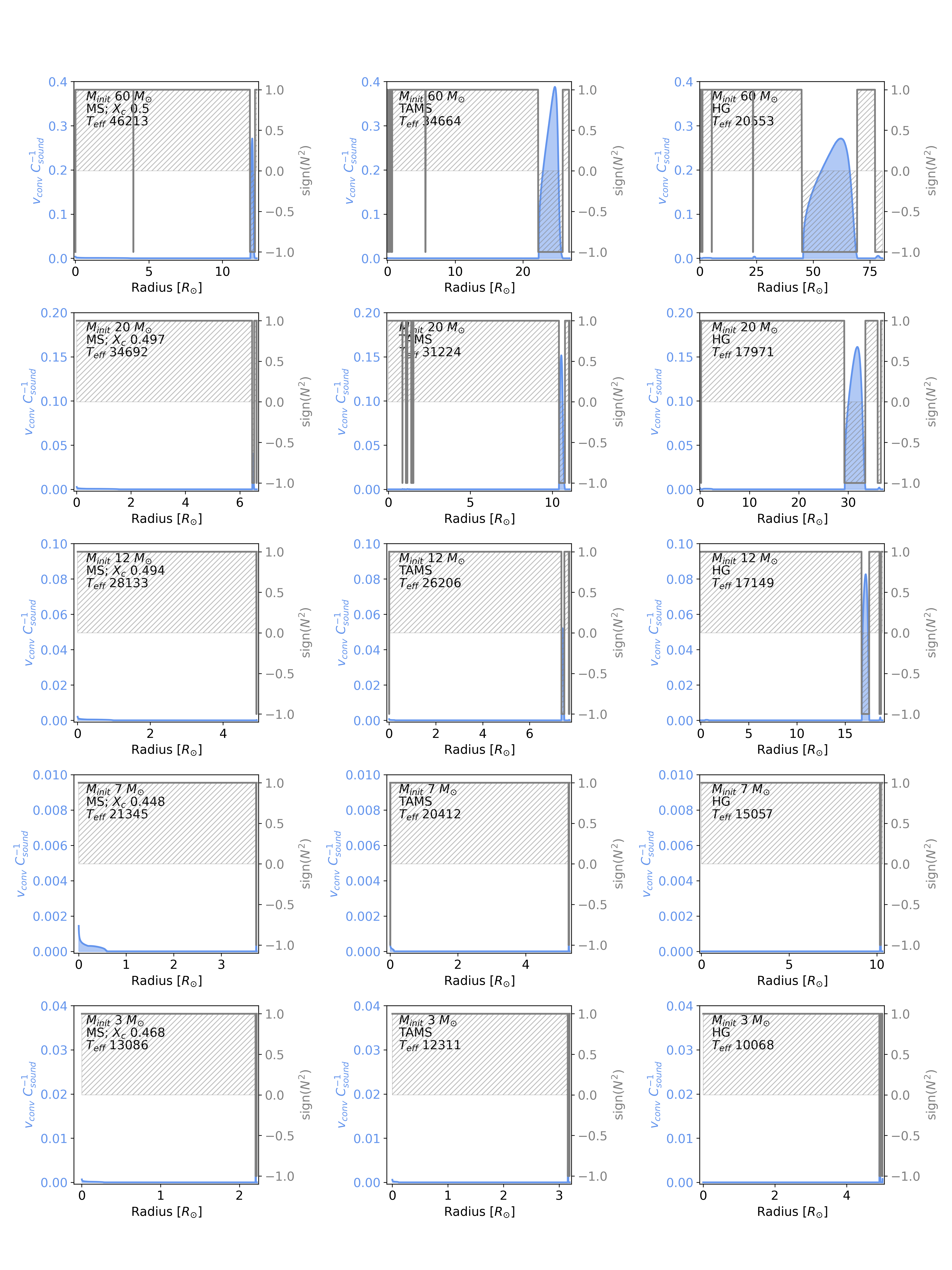}}
            
      \caption{ Profiles of the sign of the squared Brunt-Väisälä frequency (grey) and convective velocity as a fraction of sound speed (blue) as a function of radius (zoomed into outer 10\% of radius). The panels are the same as in Fig.\ref{fig:z014_opacity_convvel_vs_temperature} but for $Z$=0.006.  }
         \label{fig:prof_z006_csound_convvel_vs_radius}
   \end{centering}
\end{figure*}

\clearpage

\section{Propagation diagrams}
\label{app:propagation}

\begin{figure*}[!htbp]
   \begin{centering}
            {\includegraphics[clip,width=460pt,trim={0.0cm 0.0cm 0.0cm 0.0cm}]{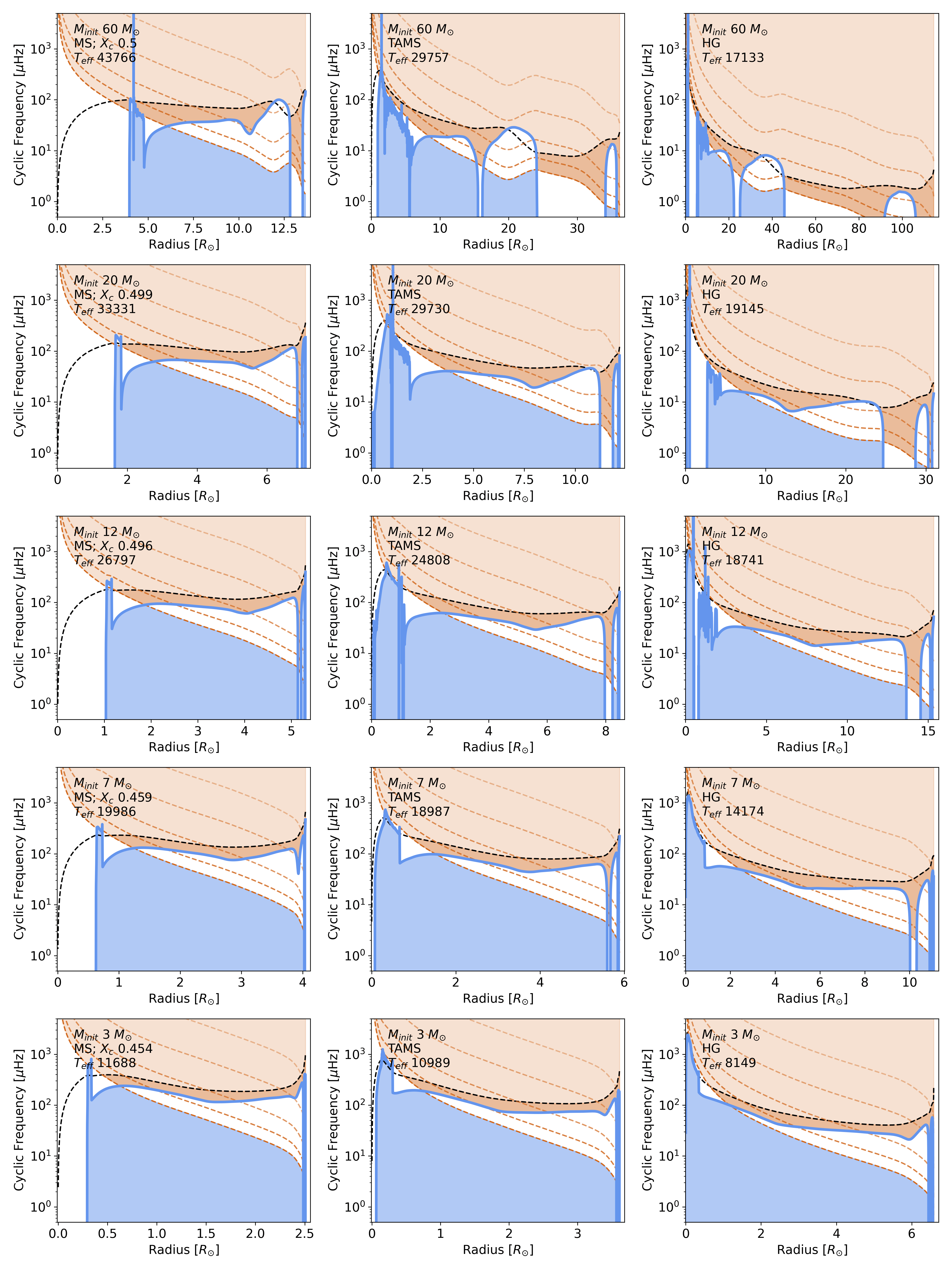}}
            
      \caption{  Propagation diagram for masses 60, 20, 12, 7, and 3 $M_{\odot}$ varying with rows. In columns, three evolutionary stages are presented: MS, TAMS, and Hertzsprung gap. $Z$=0.014. The x-axis represents the radius in solar radii, y-axis shows the cyclic frequency in microhertz. The blue line represents the Brunt-Väisälä frequency $N$, the brown dashed lines correspond to the Lamb frequencies $S_l$ for different spherical harmonic degrees $l$, and the black dashed line shows the profile of the acoustic cut-off frequency. The blue-shaded region shows the g-modes cavity, and the brown-shaded region -- the p-modes cavity.  }
         \label{fig:prof_prop_simple_z014}
   \end{centering}
\end{figure*}

\begin{figure*}
   \begin{centering}
            {\includegraphics[clip,width=480pt,trim={0.0cm 0.0cm 0.0cm 0.0cm}]{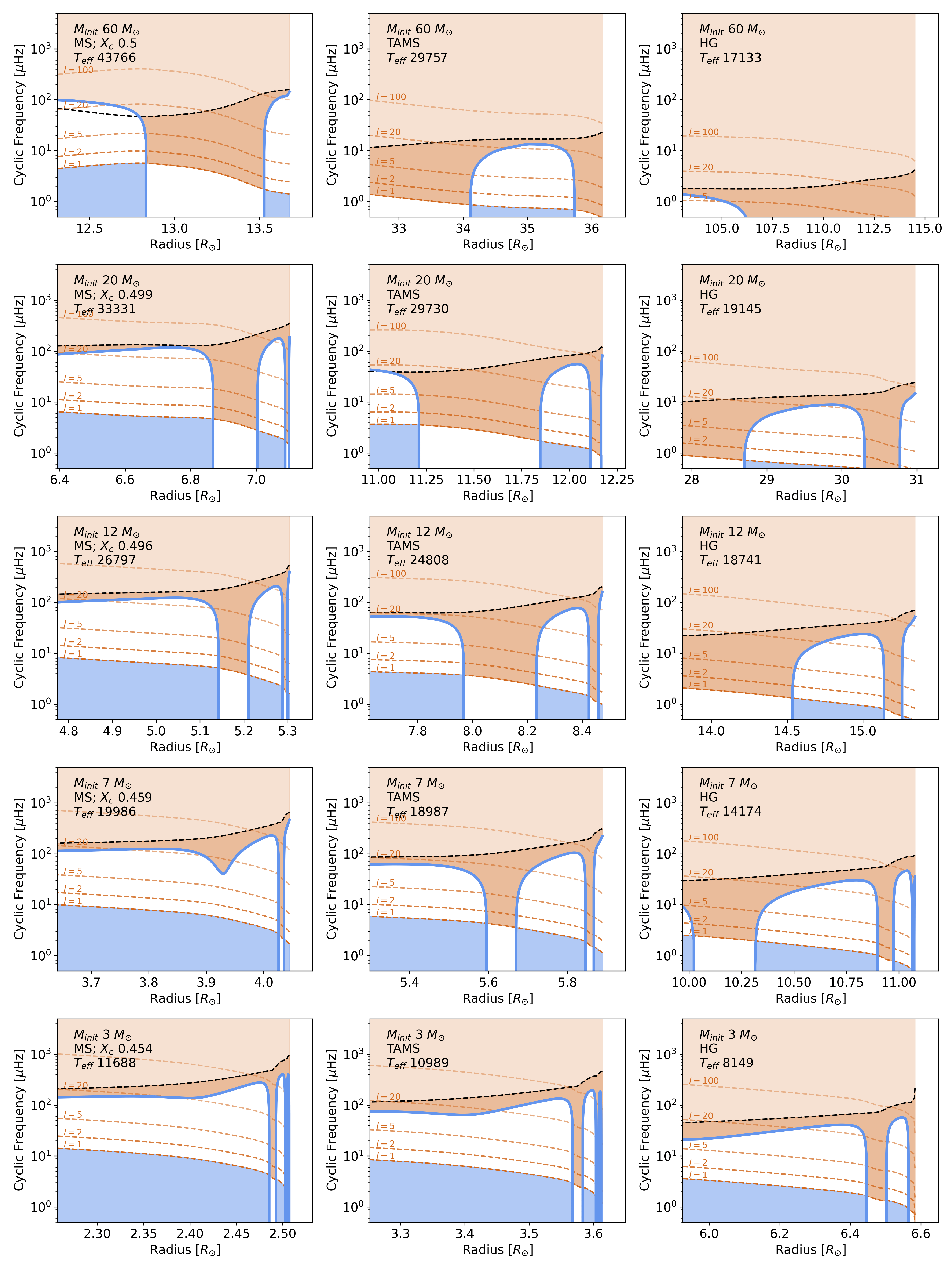}}
            
      \caption{ Same as Fig.\ref{fig:prof_prop_simple_z014}, but zoomed in external 10\% radius. }
         \label{fig:prof_prop_simple_z014_zoom}
   \end{centering}
\end{figure*}

\begin{figure*}
   \begin{centering}
            {\includegraphics[clip,width=480pt,trim={0.0cm 0.0cm 0.0cm 0.0cm}]{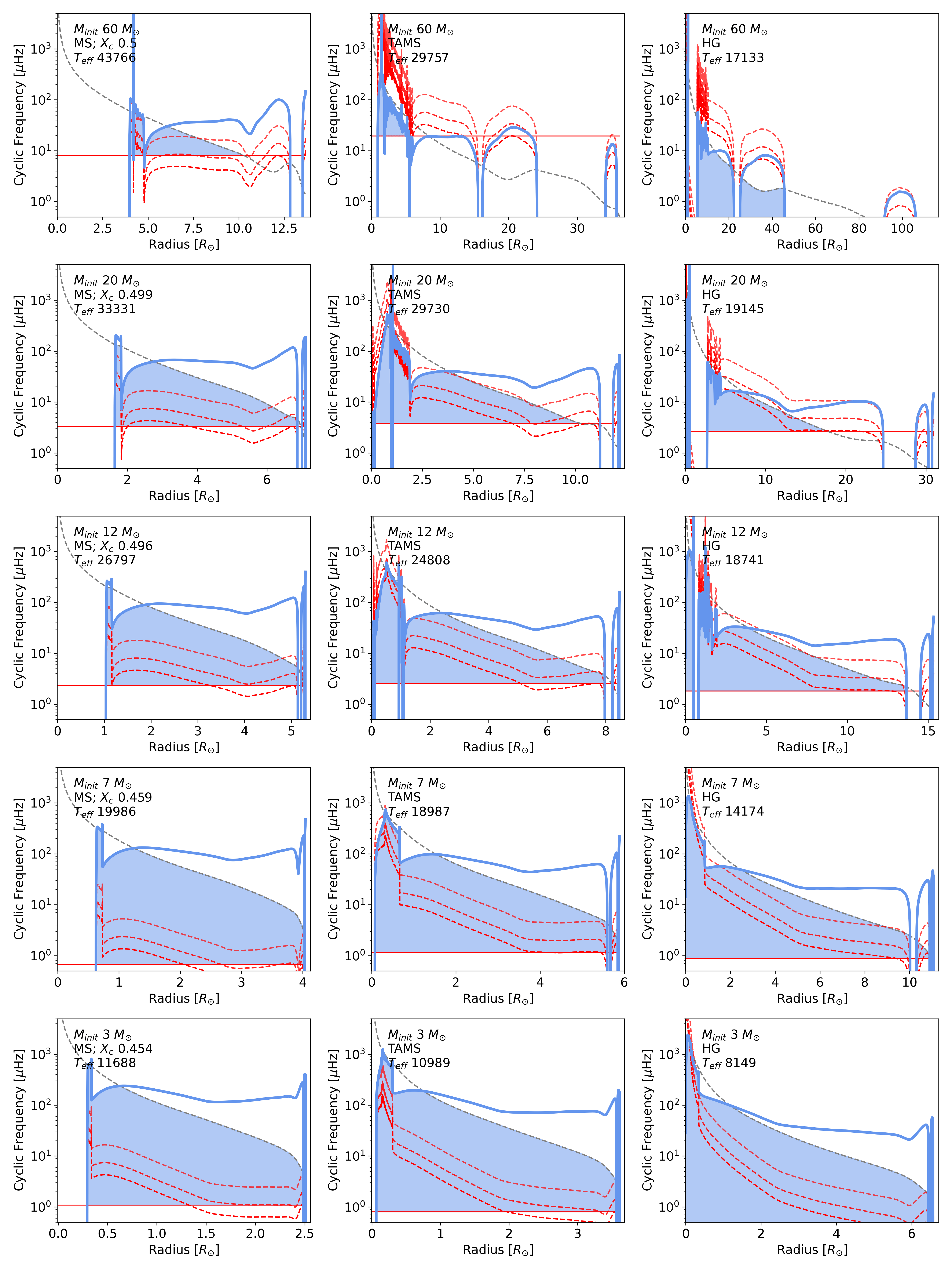}}
            
      \caption{ Same as Fig.\ref{fig:prof_prop_simple_z014}, but with the tunnelling frequency introduced as red cut-off line. }
         \label{fig:prof_prop_wmin_z014}
   \end{centering}
\end{figure*}


\begin{figure*}
   \begin{centering}
            {\includegraphics[clip,width=480pt,trim={0.0cm 0.0cm 0.0cm 0.0cm}]{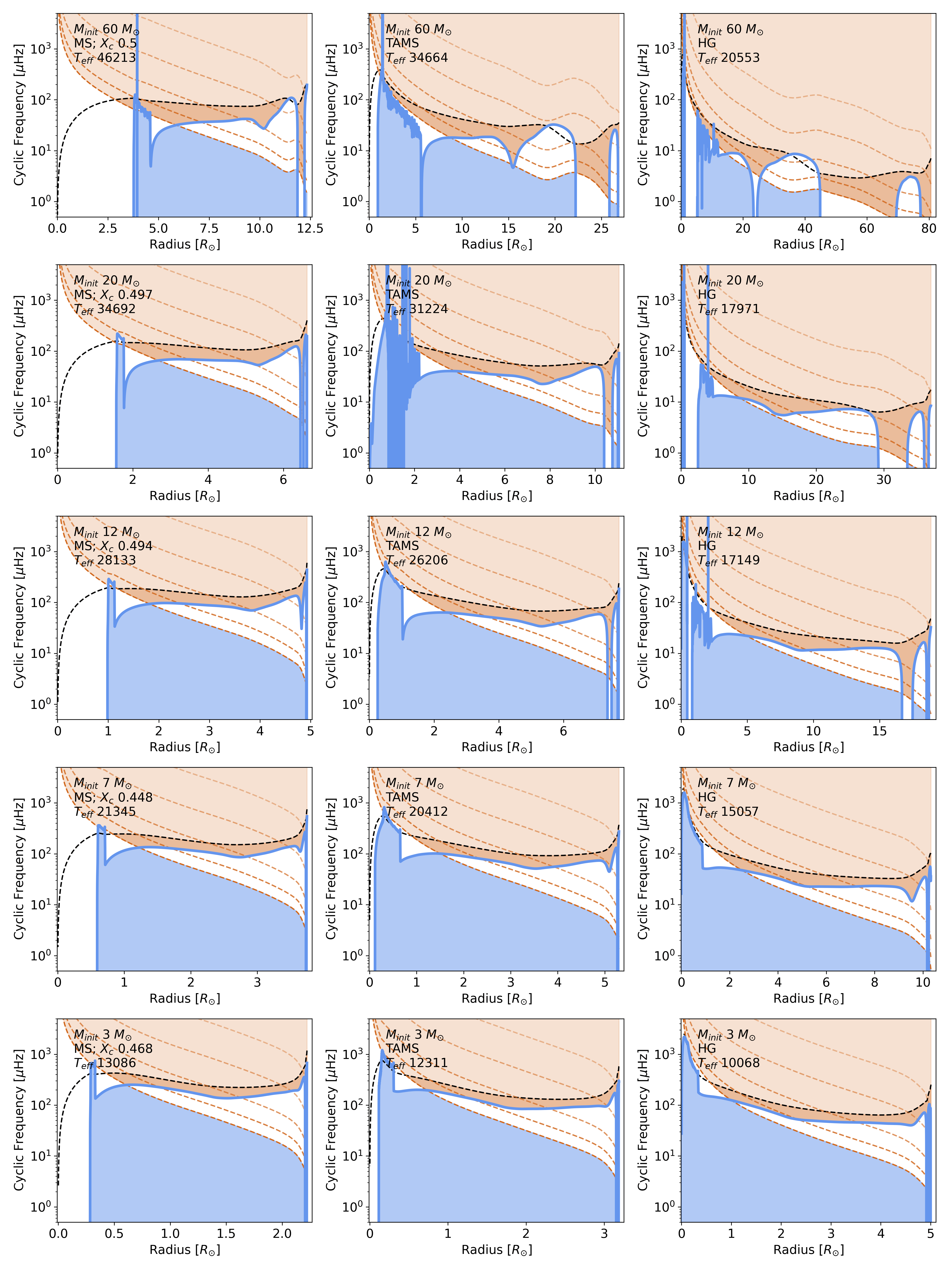}}
            
      \caption{ Same as Fig.\ref{fig:prof_prop_simple_z014}, but for $Z$=0.006. }
         \label{fig:prof_prop_simple_z006}
   \end{centering}
\end{figure*}

\begin{figure*}
   \begin{centering}
            {\includegraphics[clip,width=480pt,trim={0.0cm 0.0cm 0.0cm 0.0cm}]{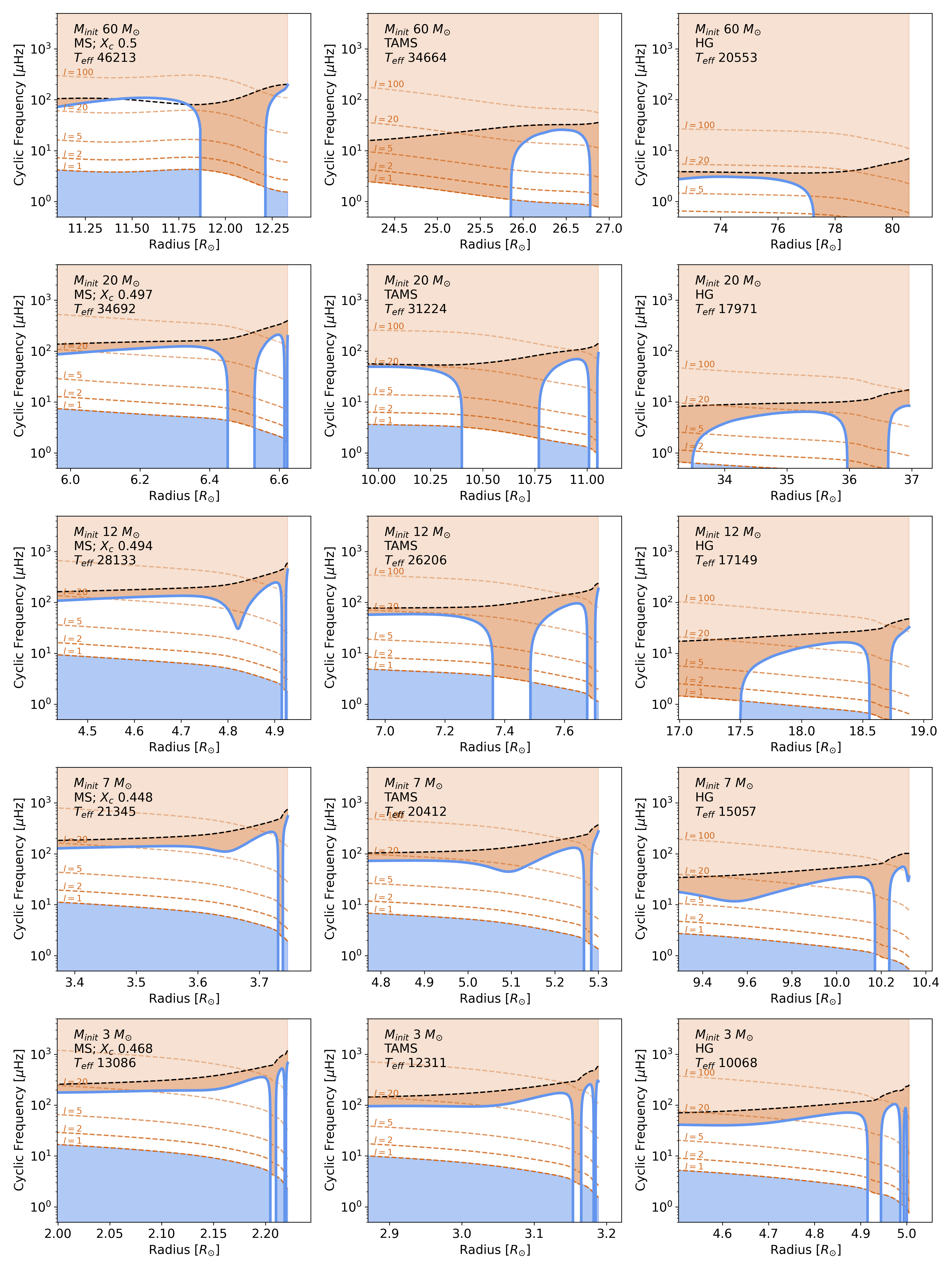}}
            
      \caption{ Same as Fig.\ref{fig:prof_prop_simple_z014}, but for $Z$=0.006 and zoomed in external 10\% radius.  }
         \label{fig:prof_prop_simple_z006_zoom}
   \end{centering}
\end{figure*}

\begin{figure*}
   \begin{centering}
            {\includegraphics[clip,width=480pt,trim={0.0cm 0.0cm 0.0cm 0.0cm}]{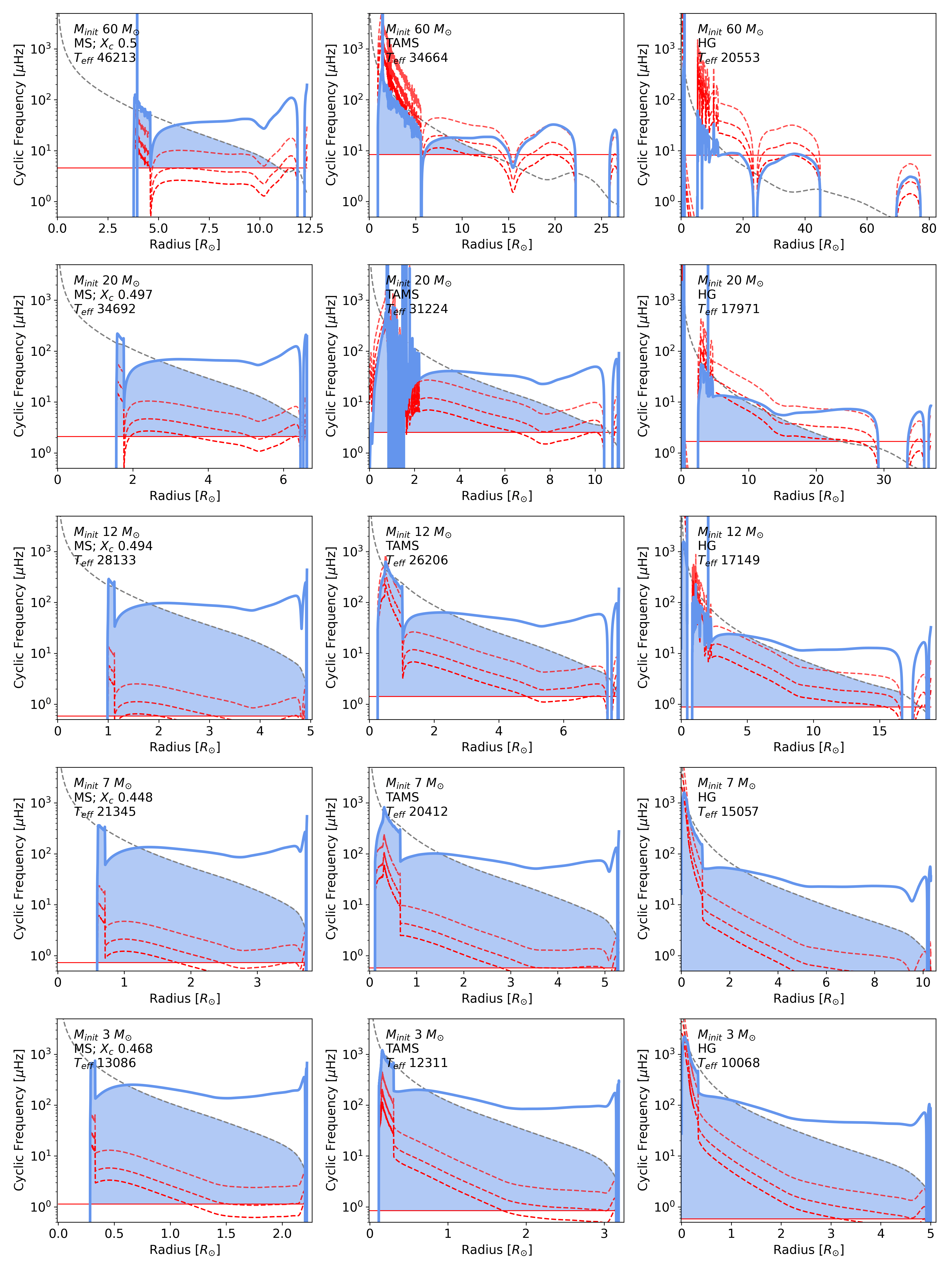}}
            
      \caption{ Same as Fig.\ref{fig:prof_prop_simple_z014}, but for $Z$=0.006 and the tunnelling frequency introduced as red cut-off line. }
         \label{fig:prof_prop_wmin_z006}
   \end{centering}
\end{figure*}



            

            

\end{document}